\RequirePackage{fix-cm}

\documentclass[smallcondensed]{svjour3}      % onecolumn (ditto)

\smartqed  % flush right qed marks, e.g. at end of proof
\usepackage{graphicx}
\usepackage{amsmath}
\usepackage{amssymb}
\usepackage{float}
\usepackage{cite}
\usepackage[colorlinks,linkcolor=blue,anchorcolor=blue,citecolor=blue]{hyperref}

\hypersetup{urlcolor=blue}
\usepackage{booktabs}
\usepackage{array}
\usepackage{makecell}
\usepackage{multirow}
\usepackage{pdflscape}
\usepackage[figuresright]{rotating}
\usepackage{rotating}
\usepackage{float}

\usepackage{tabularx}

\usepackage{ragged2e} % for \RaggedRight macro
\usepackage{booktabs} % for \toprule, \midrule etc macros
\newcolumntype{L}{>{\RaggedRight\hangafter=1\hangindent=0em}X}

\newcommand{\C}{ {\mathcal C}}

\newcommand{\bF}{ {\mathbb F}}
\newcommand{\EOP} { \hfill $\Box$ }

\newcommand{\pf} { {\rm {\bf Proof.}} }

 \journalname{}

\begin{document}
\title{Some new constructions of optimal linear codes and alphabet-optimal $(r,\delta)$-locally repairable codes %\thanks{Grants or other notes
%about the article that should go on the front page should be
%placed here. General acknowledgments should be placed at the end of the article.}
}
%\subtitle{Do you have a subtitle?\\ If so, write it here}

%\titlerunning{Short form of title}        % if too long for running head

\author{Jing Qiu  \and Fang-Wei Fu %etc.
}

%\authorrunning{Short form of author list} % if too long for running head

\institute{%\Letter~
              Jing Qiu\at
              Chern Institute of Mathematics and LPMC, Nankai University \\
              Tianjin, 300071, P. R. China\\
              \email{qiujing2077@mail.nankai.edu.cn}           %  \\
%             \emph{Present address:} of F. Author  %  if needed
             \and
             Fang-Wei Fu\at
              Chern Institute of Mathematics and LPMC, Nankai University \\
              Tianjin, 300071, P. R. China\\
              \email{fwfu@nankai.edu.cn}}

\date{Received: date / Accepted: date}

\maketitle

\begin{abstract}

In distributed storage systems, locally repairable codes (LRCs) are designed to reduce disk I/O and repair costs by enabling recovery of each code symbol from a small number of other symbols.
To handle multiple node failures, $(r,\delta)$-LRCs are introduced to enable local recovery in the event of up to $\delta-1$ failed nodes.
Constructing optimal $(r,\delta)$-LRCs has been a significant research topic over the past decade. In \cite{Luo2022}, Luo \emph{et al.}  proposed a construction of linear codes by using unions
of some projective subspaces within a projective space. Several new classes of  Griesmer codes and distance-optimal codes were constructed, and some of them were proved to be alphabet-optimal $2$-LRCs.

In this paper, we first modify the method of constructing linear codes in \cite{Luo2022} by considering a more general situation of intersecting projective subspaces. This modification enables us to construct good codes with more flexible parameters.
Additionally, we present the conditions for the constructed linear codes to qualify as Griesmer codes or achieve distance optimality. Next, we explore the locality of linear codes constructed by eliminating elements from a complete projective space. The novelty of our work lies in establishing the locality as $(2,p-2)$, $(2,p-1)$, or $(2,p)$-locality, in contrast to the previous literature that only considered $2$-locality. Moreover, by combining analysis of code parameters and the C-M like bound for $(r,\delta)$-LRCs, we construct some alphabet-optimal $(2,\delta)$-LRCs which may be either Griesmer codes or not Griesmer codes. Finally, we investigate the availability and alphabet-optimality of $(r,\delta)$-LRCs constructed from our modified framework.


% \PACS{PACS code1 \and PACS code2 \and more}
% \subclass{ 94B60 \and 11T71 }
\end{abstract}

\section{Introduction}
\label{intro}
Let $\mathbb{F}_{q}$ be the finite field with $q$ elements and $\bF_{q}^{*}=\bF_{q}\setminus \{0\}$, where $q$ is any prime power.
In this paper, we assume that $p$ is an odd prime, $m$ is a positive integer, and define $[m]\triangleq\{1,2,\dots,m\}$. Consider $\mathbb{F}_{q}^{m}$ as an $m$-dimensional vector space over $\mathbb{F}_q$, and let $\bF_{q}^{m*}=\bF_{q}^{m}\setminus \{\textbf{0}\}$, where $\textbf{0}$ denotes the zero vector.

\subsection{Griesmer codes}

A $q$-ary $[n, k, d]$ linear code $\mathcal{C}$ is a $k$-dimensional subspace of $\mathbb{F}_{q}^{n}$ with minimum distance $d$. For a $q$-ary $[n,k,d]$ linear code, the Griesmer bound is given by\cite{Griesmer1960,Solomon1965}
$$n \geq \sum_{i=0}^{k-1}\left\lceil\frac{d}{q^i}\right\rceil,$$
where $\left\lceil \cdot\right\rceil$ is the ceiling function.
A linear code achieving the Griesmer bound is called a Griesmer code. If there is no linear code with parameters $[n, k, d^{'} > d]$ for an $[n, k, d]$ linear code $\mathcal{C}$, we classify $\mathcal{C}$ as distance-optimal.

Solomon and Stiffler\cite{Solomon1965} utilized unions of mutually disjoint projective spaces to propose an infinite family of binary Griesmer codes.
More recently, Hyun \emph{et al.} \cite{Hyun2020} constructed infinite families of binary Griesmer codes by utilizing unions of projective subspaces. This construction was later generalized to the $p$-ary case by Luo \emph{et al.} \cite{Luo2022}.

\subsection{Locally repairable codes}

To reduce the repair bandwidth in massive reliable scale distributed storage system, the concept of locally repairable codes (LRCs) \cite{Gopalan2012} emerged. The $i$-th coordinate of an $[n, k]$ linear code $\mathcal{C}$ is said to have $r$-locality if the value at this coordinate can be recovered by accessing at most $r$ other coordinates. If all the coordinates have $r$-locality, we call $\mathcal{C}$ an $r$-LRC.
However, when multiple node failures occur, the original concept of locality may not work.
Prakash \emph{et al.}~\cite{Prakash2012} introduced the concept of $(r, \delta)$-locality of linear codes, where $\delta\geq 2$, which generalized the notion of $r$-locality. The $i$-th coordinate of $\C$ is said to have $(r, \delta)$-locality ($\delta \geq 2$), if there exists a subset $S_i \subset \{1, 2, \ldots, n\}$ such that $i\in S_i$, $|S_i|\leq r+\delta-1$ and the punctured code $\C|_{S_i}$ has the minimum distance $d(\C|_{S_i}) \geq \delta$, the set $S_{i}\setminus \{i\}$ is termed the repair set of $i$-th coordinate. A code $\C$ is said to have $(r, \delta)$-locality or be an $(r,\delta)$-LRC if all the coordinates of $\C$ have $(r, \delta)$-locality. Note that $(r,\delta)$-locality reduces to $r$-locality when $\delta=2$, and we call a code LRC if it has $r$-locality or $(r,\delta)$-locality. A Singleton-like bound for the minimum distance of an $(r, \delta)$-LRC is given as follows~\cite{Prakash2012}:
\begin{equation}\label{eq:singleton}
d(\C) \leq  n-k- \left(\left\lceil \frac{k}{r}\right\rceil -1 \right)(\delta-1) +1.
\end{equation}
An $(r,\delta)$-LRC achieving the bound~(\ref{eq:singleton}) is said to be Singleton-optimal.
In the last decade, many constructions of Singleton-optimal LRCs have been proposed, for example see~\cite{Tamo2014,Jin2020,Li2019,Guruswami2019,Jin2019,Chen2021it,Cai2020,Xing2019,Kong2021,Goparaju2014,Tamo2015,Tamo2016,Chen2018,Chen2019,Luo2019,Fang2018,Sun2019,Qiu2021,Fang2021} .

To consider the alphabet size and address practical application needs, Cadambe and Mazumdar \cite{Cadambe2015} introduced a new bound called the C-M bound for an $[n, k, d]$ LRC over $\mathbb{F}_q$ with locality $r$. The C-M bound is given as follows:
\begin{align}\label{eq:CM}
k\leq \min\limits_{1\leq t \leq \lceil\frac{k}{r}\rceil-1}\{tr+k_{{\rm opt}}^{(q)}(n-t(r+1), d)\},
\end{align}
where $k_{{\rm opt}}^{(q)}(n, d)$ denotes the maximum dimension of a linear code over $\mathbb{F}_q$ of length $n$ and minimum distance $d$. An $r$-LRC achieving the bound \eqref{eq:CM} is said to be alphabet-optimal.

The C-M like bound for $(r,\delta)$-LRCs was obtained in \cite{Rawat2015} as follows:
\begin{align}\label{eq:CM2}
k\leq \min\limits_{1\leq t \leq \lceil\frac{k}{r}\rceil-1}\{tr+k_{{\rm opt}}^{(q)}(n-t(r+\delta-1), d)\}.
\end{align}
An $(r,\delta)$-LRC achieving the bound \eqref{eq:CM2} is also said to be alphabet-optimal in the absence of ambiguity.

It was demonstrated in \cite{Cadambe2015} that binary Simplex codes are alphabet-optimal with locality $2$. Several infinite families of alphabet-optimal binary LRCs were proposed in \cite{Silberstein2015} by considering the punctured Simplex codes. Some alphabet-optimal binary LRCs constructed from partial spreads were presented in \cite{Ma2019}. Luo and Cao \cite{Luo2021} constructed seven infinite families of alphabet-optimal binary LRCs by using a general framework for binary linear codes. For the nonbinary cases, Silberstein and Zeh \cite{Silberstein2018} proposed several infinite families of alphabet-optimal $p$-ary LRCs with locality $2$ or $3$ by puncturing Simplex codes. Tan \emph{et al.} \cite{Tan2021} presented some infinite families of $q$-ary LRCs achieving the bound \eqref{eq:CM}, by determining the localities of some known linear codes. Very recently, Luo and Ling \cite{Luo2022} proposed more infinite families of alphabet-optimal LRCs with locality $2$ by employing the general framework of constructing $p$-ary linear codes. Note that the method in \cite{Luo2022} can be also regarded as puncturing the Simplex codes. In fact, almost all the constructed alphabet-optimal LRCs can be regarded as punctured codes of the Simplex code, and most related papers focus on the $r$-locality. In \cite{Fu2020}, Fu \emph{et al.} provided some Singleton-optimal $(r,\delta)$-LRCs from Simplex code and Cap code, but the dimensions of these codes are limited in $\{3,4\}$.

In distributed storage systems, to permit access of a coordinate from multiple ways in parallel,
LRCs was generalized to LRCs with availability in \cite{Pamies2013} and \cite{Wang2014},
in which case a coordinate has more than one repair set. For this topic, the readers may refer to \cite{Cai2020-ava},\cite{Huang2016}, \cite{Pamies2013}, \cite{Rawat2016}, \cite{Silberstein2019},\cite{Tamo2016-ava},\cite{Wang2014}.

\subsection{Our contributions and techniques}
Our contributions can be summarized as follows:

\begin{description}
\item[\textnormal{(i)}] We modify the  method of constructing linear codes proposed in \cite{Luo2022} by relaxing the restrictions on projective subspaces. This allows us to obtain some optimal codes with more flexible parameters.
\item[\textnormal{(ii)}] We provide criteria for determining the $(2, p-2)$, $(2, p-1)$, and $(2, p)$-localities of $q$-ary linear codes constructed by eliminating elements from a complete projective space. We also propose constructions for $p$-ary alphabet-optimal $(2, p-1)$, and $(2, p)$-LRCs. Notably, we prove that $p$-ary alphabet-optimal $2$-LRCs constructed in \cite{Luo2022} are also alphabet-optimal $(2,p-1)$-LRCs. Moreover, we point out that the criteria for determining the $(r,\delta)$-localities of $p$-ary codes can be generalized to determining the $(r,\delta)$-localities of $q$-ary codes, where $q$ is a prime power.
    From which, we prove that $q$-ary Simplex codes are alphabet-optimal $(2, q)$-LRCs with respect to the bound~\eqref{eq:CM2}.
\item[\textnormal{(iii)}] We demonstrate that the new linear codes constructed from the modified framework are $(r, \delta)$-LRCs with availability. Although we do not have the exact expression of the alphabet size related bound for $(r, \delta)$-LRCs with availability, we can confirm that some of new constructed codes are alphabet-optimal. Specifically, we propose a sufficient condition for these codes to be alphabet-optimal. From which, infinite families of alphabet-optimal $(r,\delta)$-LRCs with availability can be obtained. To the best of our knowledge, there has been no general construction of alphabet-optimal $(r,\delta)$-LRCs with availability.
\end{description}

This paper is organized as follows. Section 2 introduces some basic results that are needed for our discussion.  Section 3 is devoted to generalize the framework and constructions of optimal linear codes over $\mathbb{F}_p$  in \cite{Luo2022}. In Section 4, we present criteria for determining $(2,p-2)$, $(2,p-1)$, and $(2,p)$-locality of $p$-ary linear codes constructed by eliminating elements from a complete projective space. From which we can get some alphabet-optimal LRCs over $\mathbb{F}_{p}$. Note that in term of locality, the results can be generalized to $q$-ary codes without any difficulties. In Section 5, we discuss the $(r,\delta)$-locality with availability of the linear codes constructed in Section 3, a sufficient condition for these linear codes to be alphabet-optimal is provided. Finally, Section 6 concludes this paper.
%Your text comes here. Separate text sections with

\section{Preliminaries}\label{sec:pre}

\subsection{A general framework of constructing linear codes}

In this subsection, we introduce a general construction of linear codes and some basic results
about additive characters and projective spaces over finite fields.

For any vector $\textbf{x} = (x_1,\dots, x_n) \in \bF_{p}^{n}$, define the Hamming weight of $\textbf{x}$ as $\mbox{wt}(\textbf{x}) = |\{i \in [n] : x_i \neq 0\}|$. For a linear code $\mathcal{C}$, let $A_i$ denote the number of codewords in $\mathcal{C}$ with weight $i$. The sequence $(A_0, A_1, \dots, A_n)$ is called the weight distribution of $\mathcal{C}$. The weight enumerator of $\mathcal{C}$ is defined as $1 + A_1z + A_2z^2 + \dots + A_nz^n$.

In \cite{Ding2007,Ding2008}, Ding \emph{et al.} proposed a universal framework of constructing linear codes based on trace function and a nonempty subset $D = \{d_1,\dots, d_n\}\subset \bF_{p^m}$. By employing this framework, a $p$-ary linear code of length $n$ can be formed as follows: $$\mathcal{C}_D=\{\textbf{c}_{x}=(\mbox{tr}(xd_1),\dots,\mbox{tr}(xd_n)): x\in \bF_{p^m}\},$$
where $\mbox{tr}(\cdot)$ is the trace function from $\mathbb{F}_{p^m}$ to $\mathbb{F}_{p}$ given by
$$\mbox{tr}(y)=y+y^p+\dots+y^{p^{m-1}}.$$
Here, $\textbf{c}_{x}$ represents the codeword corresponding to the element $x$ in the finite field $\bF_{p^m}$. The subset $D$ is referred to as the defining set of the linear code $\mathcal{C}_D$.

 Assume that $\xi_p$ is a primitive $p$-th root of complex unity. For any $a \in \mathbb{F}_{p^m}$, an additive character of $\mathbb{F}_{p^m}$ is defined as the function $\chi_a(x) = \xi_p^{{\rm tr}(ax)}$, where $x \in \mathbb{F}_{p^m}$. All additive characters of $\mathbb{F}_{p^m}$ form a group of order $p^m$ with operation  $\chi_{a+b}(x)=\chi_a(x)\chi_b(x)$.
The famous orthogonal relation of additive characters is given as follows:
 $$\sum\limits_{x\in \bF_{p^m}}\chi_a(x)=\left\{\begin{array}{lcl}
 0,& &  \mbox{if $a\neq 0$}, \\
 p^m,& &  \mbox{if $a=0$}.
  \end{array} \right.$$

Suppose that $\{\alpha_1, \dots, \alpha_m\}$ is a basis of $\bF_{p^m}$ over $\bF_p$, then there exists a unique basis $\{\beta_1, \dots, \beta_m\}$ of $\bF_{p^m}$ over $\bF_p$ satisfying
\[
{\rm tr}(\alpha_{i}\beta_j)=\left\{\begin{array}{lcl}
 1,& &  \mbox{if $i=j$}, \\
 0,& &  \mbox{if $i\neq j$},
  \end{array} \right.
  \]
for any $1 \leq i , j \leq m$. We call $\{\beta_1, \dots, \beta_m\}$ the dual basis of $\{\alpha_1, \dots, \alpha_m\}$.
For any $x, y \in \bF_{p^m}$, we can represent them by $x = \sum^{m}_{i=1} x_i\alpha_i$ and $y =\sum^{m}_{i=1} y_i\beta_i$, where $x_i , y_i \in \bF_p$ for any $i \in [m]$. Then,
$x$ and $y$ can be expressed as vectors $\mathbf{x} = (x_1, x_2, \dots , x_m)$ and $\mathbf{y} = (y_1, y_2, \dots , y_m)$ in $\bF^m_p$, respectively.
The Euclidean inner product of $\mathbf{x}$ and $\mathbf{y}$ is defined as $\mathbf{x} \cdot \mathbf{y} = \sum^{m}_{i=1}x_i y_i$.
 It can be easily verified that $\mbox{tr}(xy) = \mathbf{x}\cdot \mathbf{y}$. Consequently, we can express the additive character $\chi_{y}(x)$ of $\bF_{p^m}$ as $\chi_{y}(x) = \xi_p^{\mathbf{x} \cdot \mathbf{y}}$.

Based on the above discussion, the previously defined linear code $\mathcal{C}_D$ is equivalent to
\begin{equation}\label{eq:framework}
\mathcal{C}_{\mathcal{D}}=\{c_{\mathbf{x}}=(\mathbf{x}\cdot \mathbf{d}_{1},\dots,\mathbf{x}\cdot \mathbf{d}_n): \mathbf{x}\in \bF_{p}^{m}\},
\end{equation}
where $\mathcal{D} = \{\mathbf{d}_1, \dots , \mathbf{d}_n\} \subset \bF^m_p$ is referred to as the defining set of $\mathcal{C}_{\mathcal{D}}$. The matrix $G=[\mathbf{d}_1^{T},\mathbf{d}_2^{T},\dots,\mathbf{d}_{n}^{T}]$ can be regarded as a generator matrix of $\mathcal{C}_{\mathcal{D}}$, and the rank of $G$ is equal to the dimension of $\mathcal{C}_{\mathcal{D}}$.

In \cite{Luo2022},  Luo \emph{et al.} introduced new constructions of Griesmer codes and distance-optimal linear codes by considering the defining set as the complement of the union of certain projective subspaces within a projective space.

We will follow the notations in \cite{Luo2022}. Let $V$ be the $m$-dimensional vector space $\mathbb{F}_p^{m}$. Two nonzero vectors $\mathbf{x}=(x_1,x_2,\dots, x_m)$ and $\mathbf{y}=(y_1 , y_2 , \dots, y_m)$ in $V$ are said to be equivalent, denoted by $\textbf{x}\sim \textbf{y}$, if $\mathbf{y} = \lambda\mathbf{x}$ for some $\lambda$ in $\mathbb{F}_p^{*}$. The relation $\sim$ is indeed an equivalence relation. Denote $(x_1 : x_2 : \dots: x_m)$ the equivalent class consists of all nonzero scalar multiples of $(x_1 , x_2 , \dots, x_m)$.
The set of all equivalent classes in $V$ is a projective space over $\mathbb{F}_p$ with dimension $m-1$, termed the projective space of $V$. The elements of a projective space are called points. For every point $(x_1 : x_2 : \dots: x_m)$ in the projective space of $V$, we can use arbitrary nonzero scalar multiple of $(x_1 , x_2 , \dots, x_m)$ to express the point.

Let $\mathcal{A}$ be a nonempty subset of $[m]$. Define an $|\mathcal{A}|$-dimensional vector space over $\bF_p$ by
\begin{equation}\label{eq:LA}
L_{\mathcal{A}}=\{(a_1, \dots , a_n):a_i\in \bF_p {\rm ~if~} i\in \mathcal{A} {\rm ~and~} a_i=0 {\rm ~if~}i\notin  \mathcal{A}\}.
\end{equation}
Assume that $P_{\mathcal{A}}$ is the projective space of $L_{\mathcal{A}}$.
For convenience, we assign the expression of every point in $P_{\mathcal{A}}$ as the vector of $\bF_p^{m}$ in the corresponding equivalent class whose first nonzero coordinate is $1$. In this way, $P_{\mathcal{A}}$ can be regarded as a subset of $L_{\mathcal{A}}$.

It is easy to check that $|P_{\mathcal{A}}| = \frac{p^{|\mathcal{A}|}-1}{p-1}$ and
$L_{\mathcal{A}}\setminus \{0\} = \bigcup_{a\in \bF_{p}^{*}}aP_{\mathcal{A}}$.
Obviously,
$L_{\mathcal{A}}\setminus \{0\} = P_{\mathcal{A}}$ if $p = 2$. For any two subsets $\mathcal{A}_1,\mathcal{A}_2$ of $[m]$, the intersection of $P_{\mathcal{A}_1}$ and
$P_{\mathcal{A}_2}$ is equal to $P_{\mathcal{A}_1\cap \mathcal{A}_2}$, where $P_{\emptyset} = \emptyset$.

\subsection{Modifying the framework}

As we can see from \eqref{eq:framework}, the defining set in the original framework is required to be a subset of $\bF^m_p$. In this subsection, we modify the framework by allowing defining set to be a multi-set consisting of vectors from $\bF^m_p$.

\begin{definition}{ (Modified framework)}
Suppose that $s$ is a positive integer. Let $\mathcal{D}_1, \mathcal{D}_2,\dots, \mathcal{D}_{s}$ be subsets of $\bF^m_p$.
We can define a $p$-ary linear code by
\begin{equation}\label{eq:define}
\mathcal{C}_{(\mathcal{D}_1,\mathcal{D}_2,\dots,\mathcal{D}_s)}=\{c_{\mathbf{x}}=(\mathbf{x} G_1,\dots,\mathbf{x} G_s): x\in \bF_{p}^{m}\},
\end{equation}
where $G_i$ denotes the matrix whose columns are transpose of vectors of $\mathcal{D}_i$ for all $i\in [s]$.
\end{definition}

Following the approach outlined in \cite{Luo2022}, we can consider each $\mathcal{D}_i$($1\leq i
\leq s$) as the complement of unions of specific projective subspaces within a projective space. Consequently, the construction of good linear codes can be simplified to design suitable subsets of $[m]$.

\begin{definition}
Suppose that $t > 1$ is an integer. Let $E=\{\mathcal{E}_1, \mathcal{E}_2,\dots, \mathcal{E}_{t}\}$ be a multi-set with elements being nonempty subsets of $[m]$, we call set
$$\bigcup_{1\leq i<j\leq t}(\mathcal{E}_i \cap \mathcal{E}_j)$$
the \textbf{center} of $E$, denoted by ${\rm Center}(E)$.

\end{definition}

\begin{definition}{ (Property I$_{s}$)}
Suppose that $\ell > 1$ is a positive integer, $\mathcal{A}_1, \mathcal{A}_2,\dots, \mathcal{A}_{\ell}$ are nonempty subsets of $[m]$. If we can partition the multi-set  $A=\{\mathcal{A}_{i}\}_{i=1}^{\ell}$ into the form as $A=B_{1}\cup B_{2}\cup \cdots \cup B_{s},$
%$$\{\mathcal{A}_{i}\}_{i=1}^{\ell}=\bigcup_{j=1}^{s}B_{j},$$
where
\begin{equation}\label{eq:B}
B_{j}=\{\mathcal{B}_{1}^{(j)}, \mathcal{B}_{2}^{(j)},\dots, \mathcal{B}_{\ell_{j}}^{(j)}\} {\rm \,\,for\,\, any\,\,} 1\leq j\leq s,
\end{equation}
$s$, $\ell_1,\ell_2,\dots,\ell_s$ are positive integers satisfying $\ell=\sum_{i=1}^{s}\ell_i$, such that
$$\mathcal{A}_{i} \setminus \bigcup_{j=1}^{s}  {\rm Center}(B_{j})\neq \emptyset$$ for any $1\leq i \leq \ell$,
then  $\mathcal{A}_1, \mathcal{A}_2,\dots, \mathcal{A}_{\ell}$ are said to satisfy Property I$_{s}$.
\end{definition}

In \cite{Luo2022}, the authors initially required that $\mathcal{A}_1, \mathcal{A}_2,\dots, \mathcal{A}_{\ell}$ are nonempty subsets of $[m]$ satisfying $\mathcal{A}_i \setminus \bigcup_{j\in [\ell]\setminus \{i\}}\mathcal{A}_{j}\neq \emptyset$ for every $i \in [\ell]$. These requirements are in fact equivalent to Property I$_{s}$ with $s=1$ by the following lemma.

\begin{lemma}
Suppose that $\ell > 1$ is a positive integer. Let $\mathcal{A}_1, \mathcal{A}_2,\dots, \mathcal{A}_{\ell}$ be nonempty subsets of $[m]$, and let $A$ be the multi-set $\{\mathcal{A}_{i}\}_{i=1}^{\ell}$. For any $i\in [\ell]$, $\mathcal{A}_i \setminus \bigcup_{j\in [\ell]\setminus \{i\}}\mathcal{A}_{j}\neq \emptyset$ if and only if $\mathcal{A}_i\setminus {\rm Center}(A)\neq \emptyset$.
\end{lemma}
\pf
 For any $i\in [\ell]$, we have

\begin{align*}
\mathcal{A}_{i}\setminus {\rm Center}(A)&= \mathcal{A}_{i}\setminus \bigcup_{1\leq j<k\leq \ell}(\mathcal{A}_j \cap \mathcal{A}_k)\\
&=  \mathcal{A}_{i}\setminus \left(\left(\mathcal{A}_{i}\cap \bigcup_{j\in [\ell]\setminus \{i\}}\mathcal{A}_{j}\right)\cup {\rm Center}(A\setminus \{\mathcal{A}_{i}\})  \right)\\
&=  \mathcal{A}_{i}\setminus \left(\bigcup_{j\in [\ell]\setminus \{i\}}\mathcal{A}_{j} \cup {\rm Center}(A\setminus \{\mathcal{A}_{i}\})  \right)\\
&=\mathcal{A}_{i}\setminus \bigcup_{j\in [\ell]\setminus \{i\}}\mathcal{A}_{j},
\end{align*}
where the last equation comes from ${\rm Center}(A\setminus \{\mathcal{A}_{i}\}) \subset \bigcup_{j\in [\ell]\setminus \{i\}}\mathcal{A}_{j}$.

The proof is completed.

\EOP

\subsection{Some auxiliary lemmas}

For $\mathcal{D} \subset \bF^m_p$, $\mathcal{A} \subset [m]$ and $\mathbf{x} \in \bF^m_p$, let $\chi_{\mathbf{x}}(\mathcal{D}) = \sum_{\mathbf{y}\in \mathcal{D} }\xi_p^{\mathbf{x}\cdot \mathbf{y}}$ and let $\mathbf{x}_{\mathcal{A}}$ be a vector obtained from $\mathbf{x}$ by removing the coordinates in $[m] \setminus \mathcal{A}$.

The following two lemmas play a fundamental role in \cite{Luo2022} and are also relevant to our proofs.

\begin{lemma}\emph{\cite[Lemma~2.1]{Luo2022}}\label{lem:lianggejihe}
Assume that $\mathcal{A}_1$ and $\mathcal{A}_2$ are subsets of $[m]$ such that they do not contain each
other. Let $P_{\mathcal{A}_i}$ be the projective space of $L_{\mathcal{A}_i}$ defined as in \eqref{eq:LA}, $i = 1, 2$. Then, for any $\mathbf{x} \in \mathbb{F}_{p}^{m*}$, we have
\begin{equation*}
{\rm wt}(c_{\mathbf{x}}) =\left\{
\begin{array}{lcl}
p^{|\mathcal{A}_1|}+P^{|\mathcal{A}_2|}-P^{|\mathcal{A}_1\cap \mathcal{A}_2|}-1,& &  {\rm if }\,\, \mathbf{x}_{\mathcal{A}_1}=\mathbf{0},\mathbf{x}_{\mathcal{A}_2}=\mathbf{0}, \\
p^{|\mathcal{A}_1|}-P^{|\mathcal{A}_1\cap \mathcal{A}_2|}-1,& &  {\rm if }\,\, \mathbf{x}_{\mathcal{A}_1}=\mathbf{0},\mathbf{x}_{\mathcal{A}_2}\neq \mathbf{0},\\
p^{|\mathcal{A}_2|}-P^{|\mathcal{A}_1\cap \mathcal{A}_2|}-1,& &  {\rm if }\,\,
\mathbf{x}_{\mathcal{A}_1}\neq\mathbf{0},\mathbf{x}_{\mathcal{A}_2}=\mathbf{0}, \\
-p^{|\mathcal{A}_1\cap \mathcal{A}_2|}-1,& &  {\rm if }\,\,
\mathbf{x}_{\mathcal{A}_1}\neq\mathbf{0},\mathbf{x}_{\mathcal{A}_2}\neq\mathbf{0}, \mathbf{x}_{\mathcal{A}_1\cap\mathcal{A}_2}=\mathbf{0}, \\
-1,& &  {\rm if }\,\,
\mathbf{x}_{\mathcal{A}_1}\neq\mathbf{0},\mathbf{x}_{\mathcal{A}_2}\neq\mathbf{0}, \mathbf{x}_{\mathcal{A}_1\cap\mathcal{A}_2}\neq\mathbf{0}.
 \end{array} \right.
\end{equation*}

\end{lemma}

\begin{lemma}\emph{\cite[Lemma~2.2]{Luo2022}}\label{lem:jixiaojuli}
Suppose that $\ell > 1$ is a positive integer. Let $\mathcal{A}_1,\mathcal{A}_2,$ $\dots ,\mathcal{A}_{\ell}$ be nonempty subsets
of $[m]$ satisfying $\mathcal{A}_i \setminus \bigcup_{j\in [\ell]\setminus \{i\}}\mathcal{A}_{j}\neq \emptyset$ for any $i \in [\ell]$. Then
\[
\min \left\{
  \sum_{y\in \bF_{p}^{*}}\chi_{\mathbf{x}}\left(y\left(\bigcup_{i=1}^{\ell}P_{\mathcal{A}_i}\right)\right):\mathbf{x}\in \bF_{p}^{m*}
\right\}=-1+\sum_{k=2}^{\ell}(-1)^{k-1}\sum_{1\leq i_1<\dots<i_k\leq \ell}p^{|\bigcap_{j=1}^{k}\mathcal{A}_{i_j}|}
.\]
\end{lemma}

In \cite{Luo2022}, the authors also pointed out that
$$\sum_{y\in \bF_{p}^{*}}\chi_{\mathbf{x}}\left(y\left(\bigcup_{i=1}^{\ell}P_{\mathcal{A}_i}\right)\right)
=-1+\sum_{k=2}^{\ell}(-1)^{k-1}\sum_{1\leq i_1<\dots<i_k\leq \ell}p^{|\bigcap_{j=1}^{k}\mathcal{A}_{i_j}|}$$
if and only if $\mathbf{x}_{\mathcal{A}_i} \neq \mathbf{0}$ for all $i \in [\ell]$ and $\mathbf{x}_{\mathcal{A}_{i_1}\cap \mathcal{A}_{i_{2}}} = \mathbf{0}$ for all $1 \leq i_1 < i_2 \leq \ell$.\\

Next, we generalize Lemma~\ref{lem:jixiaojuli} to the case of $s\geq 1$.

\begin{lemma}\label{lem:char}
Suppose that $\ell > 1$, $s$ are positive integers. If $\mathcal{A}_1, \mathcal{A}_2,\dots, \mathcal{A}_{\ell}$ are nonempty subsets of $[m]$ satisfying Property {\rm I}$_{s}$, let $B_{j}=\{\mathcal{B}_{i}^{(j)}\}_{i=1}^{\ell_j}$ for all $1\leq j\leq s$ are defined as in \eqref{eq:B}, then
\begin{align*}
& \min \left\{ \sum_{j=1}^{s}\sum_{y\in \bF_{p}^{*}}\chi_{\mathbf{x}}\left(y\left(\bigcup_{i=1}^{\ell_j}P_{\mathcal{B}_i^{(j)}}\right)\right):\mathbf{x} \in \bF_{p}^{m*} \right\} \\
=&\sum_{r=1}^{s}\left(-1+\sum_{k=2}^{\ell_{r}}(-1)^{k-1}\sum_{1\leq i_1<\dots<i_k\leq \ell_{r}}p^{|\bigcap_{j=1}^{k}\mathcal{B}_{i_j}^{(r)}|}\right).
\end{align*}
\end{lemma}

\pf
From Lemma~\ref{lem:jixiaojuli} we know
\begin{align*}
& \min \left\{ \sum_{j=1}^{s}\sum_{y\in \bF_{p}^{*}}\chi_{\mathbf{x}}\left(y\left(\bigcup_{i=1}^{\ell_j}P_{\mathcal{B}_i^{(j)}}\right)\right):\mathbf{x} \in \bF_{p}^{m*} \right\} \\
=&\sum_{r=1}^{s}\left(-1+\sum_{k=2}^{\ell_{r}}(-1)^{k-1}\sum_{1\leq i_1<\dots<i_k\leq \ell_{r}}p^{|\bigcap_{j=1}^{k}\mathcal{B}_{i_j}^{(r)}|}\right)
\end{align*}
if and only if
$$\sum_{y\in \bF_{p}^{*}}\chi_{\mathbf{x}}\left(y\left(\bigcup_{i=1}^{\ell_r}P_{\mathcal{B}_i^{(r)}}\right)\right)=-1+\sum_{k=2}^{\ell_{r}}(-1)^{k-1}\sum_{1\leq i_1<\dots<i_k\leq \ell_{r}}p^{|\bigcap_{j=1}^{k}\mathcal{B}_{i_j}^{(r)}|}$$
for all $1\leq r\leq s$,  if and only if the following conditions are satisfied simultaneously,
\begin{description}
  \item[\textnormal{(i)}] $\mathbf{x}_{\mathcal{B}_j^{(i)}}\neq \mathbf{0}$ for any  $j\in  [\ell_{i}]$ and $i \in [s]$,
  \item[\textnormal{(ii)}] $\mathbf{x}_{\mathcal{B}_{i_1}^{(u)} \cap \mathcal{B}_{i_2}^{(u)} }= \mathbf{0}$ for any $1 \leq i_1 < i_2 \leq \ell_u$ and  $1\leq u\leq s$.
\end{description}
The above conditions are equivalent to
\begin{description}
  \item[\textnormal{(i)$^{*}$}] $\mathbf{x}_{\mathcal{A}_{i}} \neq \mathbf{0}$ for any  $i \in [\ell]$,
  \item[\textnormal{(ii)$^{*}$}] $\mathbf{x}_{\cup_{j=1}^{s}  {\rm Center}(B_{j})}= \mathbf{0}$.
\end{description}
Such $\mathbf{x}$ always exists due to Property I$_{s}$.
\EOP

\section{New constructions of optimal linear codes}

In \cite{Luo2022}, by setting the defining set to be complement of unions of some projective subspaces within $P_{[m]}$, the authors obtained some Griesmer codes and distance-optimal codes with respect to the Griesmer bound. In this section, we generalize the construction in \cite{Luo2022} by using \eqref{eq:define} and subsets of $[m]$ satisfying Property I$_{s}$. Specifically, we require each $\mathcal{D}_{i}$ to be the complement of unions of some projective subspaces within $P_{[m]}$.

\begin{theorem}\label{thm:construction}
Let $p$ be an odd prime,  $m$, $\ell > 1$ and $s$ be positive integers.
Suppose that $\mathcal{A}_1, \mathcal{A}_2,\dots, \mathcal{A}_{\ell}$ are nonempty subsets of $[m]$ satisfying Property {\rm I}$_{s}$, and $B_{j}=\{\mathcal{B}_{i}^{(j)}\}_{i=1}^{\ell_i}$ for all $j
\in [s]$ are defined as in \eqref{eq:B}. Let $\mathcal{D}_i=P_{[m]}\setminus \bigcup_{j=1}^{\ell_{i}} P_{\mathcal{B}_{j}^{(i)}}$,  $\mathcal{D}_i^{c}=\bigcup_{j=1}^{\ell_{i}} P_{\mathcal{B}_{j}^{(i)}}$ for every $i
\in [s]$.
If $p^{m-1}>\sum_{i=1}^{\ell_{r}}p^{|\mathcal{B}_{i}^{(r)}|-1}$ for all $r
\in [s]$,
then $\mathcal{C}_{(\mathcal{D}_1,\mathcal{D}_2,\dots,\mathcal{D}_s)}$ defined by \eqref{eq:define} is a linear code over $\bF_p$ with parameters
$$\left[s\frac{p^m-1}{p-1}-\sum_{r=1}^{s}|\mathcal{D}_{r}^{c}|,m,sp^{m-1}-\sum_{i=1}^{\ell}p^{|\mathcal{A}_i|-1}\right],$$
where
\[
|\mathcal{D}_{r}^{c}|=\frac{\sum_{k=1}^{\ell_{r}}(-1)^{k-1}\sum_{1\leq i_1<\dots<i_k\leq \ell_{r}}p^{|\bigcap_{j=1}^{k}\mathcal{B}_{i_j}^{(r)}|}-1}{p-1}, r=1,\dots,s.
\]
\end{theorem}

\pf
 From the principle of inclusion and exclusion (PIE), we get the form of $|\mathcal{D}_{r}^{c}|$ for each $r \in [s]$ directly.

For any $\mathbf{x} \in \bF_{p}^{m*}$, by the orthogonal relation of additive characters, we have
\begin{align*}
{\rm wt}(c_{\mathbf{x}}) &= \sum_{r=1}^{s}\left(\frac{p^m-1}{p-1}-|\mathcal{D}_{r}^{c}|-|\{\mathbf{d}\in \mathcal{D}_{r}: \mathbf{x}\cdot \mathbf{d}=0\}|\right)\\
  &\overset{(a)}{=}\sum_{r=1}^{s}\left(\left(\frac{p^m-1}{p-1}-|\mathcal{D}_{r}^{c}|\right)\frac{p-1}{p}+\frac{1}{p}+\frac{1}{p}\sum_{y\in \bF_{p}^{*}}\chi_{\mathbf{x}}\left(y\left(\bigcup_{i=1}^{\ell_{r}}P_{\mathcal{B}_{i}^{(r)}}\right)\right)\right),
\end{align*}
where $(a)$ can be found in the proof of Theorem 3.1 of \cite{Luo2022}.

It then follows from Lemma~\ref{lem:char} that the minimum value of ${\rm wt}(c_{\mathbf{x}})$ for any $\mathbf{x} \in \bF_{p}^{m*}$ is
$$\sum_{r=1}^{s}\left(p^{m-1}- \sum_{i=1}^{\ell_{r}}p^{|\mathcal{B}_{i}^{(r)}|-1}\right)=sp^{m-1}-\sum_{i=1}^{\ell}p^{|\mathcal{A}_i|-1}.$$

So the minimum distance of $\mathcal{C}_{(\mathcal{D}_1,\mathcal{D}_2,\dots,\mathcal{D}_s)}$ is $sp^{m-1} -\sum_{i=1}^{\ell}p^{|\mathcal{A}_i|-1}>0$.
It is evident that ${\rm wt}(c_{\mathbf{x}})= 0$  if and only if $\mathbf{x} = \mathbf{0}$, hence $\mathcal{C}_{(\mathcal{D}_1,\mathcal{D}_2,\dots,\mathcal{D}_s)}$ has dimension $m$.

\EOP

Next, we will discuss the optimality of the linear codes given by Theorem~\ref{thm:construction} with respect to the
Griesmer bound. We follow the notations in \cite{Luo2022}.
Let $\mathcal{A}_{1}, \mathcal{A}_{2},\dots, \mathcal{A}_{\ell}$ be subsets of $[m]$.
Assume that $|\mathcal{A}_{1}| = \dots = |\mathcal{A}_{i_{1}}| =s_1, |\mathcal{A}_{i_{1}+1}| = \dots= |\mathcal{A}_{i_{2}}| = s_2,\dots, |\mathcal{A}_{i_{t-1}+1}| = \dots = |\mathcal{A}_{\ell}| = s_{t}$, where $s_1 < s_2 < \dots < s_t$ and $t \leq \ell$.
Then $\sum_{i=1}^{\ell} p^{|\mathcal{A}_{i}|} = \sum_{i=1}^{t}a_{i}p^{s_{i}}$, where $a_i$ denotes the number of subsets of size $s_i$ in $\mathcal{A}_{1}, \mathcal{A}_{2},\dots, \mathcal{A}_{\ell}$ for any $i \in [t]$.
Put $M(\mathcal{A}_{1}, \mathcal{A}_{2},\dots, \mathcal{A}_{\ell}) = \max\{a_i : i = 1,\dots, t\}$. Suppose that $P(\sum_{i=1}^{\ell} p^{|\mathcal{A}_{i}|-1}) =\sum_{i=g}^{h}b_{i}p^{i}$ is the $p$-adic expansion of $\sum_{i=1}^{\ell} p^{|\mathcal{A}_{i}|-1}$ with coefficients $b_i$ in $\{0, 1,\dots, p-1\}$ and $b_g\neq 0$, $b_h \neq 0$.
Let $C(\sum_{i=1}^{\ell}p^{|\mathcal{A}_i|-1})=\sum_{i=g}^{h}b_i$ and let $v_p(\sum_{i=1}^{\ell}p^{|\mathcal{A}_i|-1})$ be the
$p$-adic valuation of $\sum_{i=1}^{\ell}p^{|\mathcal{A}_i|-1}$.
It is easy to see that $v_p(\sum_{i=1}^{\ell}p^{|\mathcal{A}_i|-1})= g$.

\begin{theorem}\label{thm:iff}
Let the notation be the same as in Theorem~\ref{thm:construction}.
\begin{itemize}
  \item[\textnormal{(1)}]Then $\mathcal{C}_{(\mathcal{D}_1,\mathcal{D}_2,\dots,\mathcal{D}_s)}$ defined by \eqref{eq:define} is a Griesmer code if and only if $\mathcal{B}_{1}^{(r)},\mathcal{B}_{2}^{(r)}, \\ \dots ,\mathcal{B}_{\ell_{r}}^{(r)}$ are mutually disjoint for each $r\in [s]$ and
$$M(\mathcal{A}_1,\mathcal{A}_2, \dots ,\mathcal{A}_{\ell}) \leq p- 1.$$
  \item[\textnormal{(2)}]If $$\sum_{r=1}^{s}|\mathcal{D}_{r}^{c}|> \frac{\sum_{i=1}^{\ell} p^{|\mathcal{A}_i|}-C(\sum_{i=1}^{\ell} p^{|\mathcal{A}_i|-1})}{p-1}-v_p(\sum_{i=1}^{\ell}p^{|\mathcal{A}_i|-1})-1,$$ then $\mathcal{C}_{(\mathcal{D}_1,\mathcal{D}_2,\dots,\mathcal{D}_s)}$ is distance-optimal with respect to the Griesmer bound.
\end{itemize}
\end{theorem}

\pf
(1)~Note that $\mathcal{C}_{(\mathcal{D}_1,\mathcal{D}_2,\dots,\mathcal{D}_s)}$ is a $p$-ary linear code with parameters $$\left[s\frac{p^m-1}{p-1}-\sum_{r=1}^{s}|\mathcal{D}_{r}^{c}|,m,sp^{m-1}-\sum_{i=1}^{\ell}p^{|\mathcal{A}_i|-1}\right].$$ From
\begin{align}
\sum_{j=0}^{m-1}\left\lceil\frac{sp^{m-1}-\sum_{i=1}^{\ell}p^{|\mathcal{A}_i|-1}}{p^j}\right\rceil &= \sum_{j=0}^{m-1}\left\lceil\frac{sp^{m-1}-P(\sum_{i=1}^{\ell}p^{|\mathcal{A}_i|-1})}{p^j}\right\rceil  \nonumber\\
&=\sum_{j=0}^{m-1}\left\lceil\frac{sp^{m-1}-\sum_{i=g}^{h}b_ip^i}{p^j}\right\rceil \nonumber \\
&=s\sum_{j=0}^{m-1}p^{m-1-j}-\sum_{i=g}^{h}b_i\left(\sum_{j=0}^{i}p^{i-j}\right) \nonumber\\
&=s\frac{p^m-1}{p-1}-\frac{\sum_{i=1}^{\ell}p^{|\mathcal{A}_i|}-C(\sum_{i=1}^{\ell}p^{|\mathcal{A}_i|-1})}{p-1}\label{eq:616}
,\end{align}
we know that $\mathcal{C}_{(\mathcal{D}_1,\mathcal{D}_2,\dots,\mathcal{D}_s)}$ is a Griesmer code if and only if

\begin{align*}
&\frac{\sum_{i=1}^{\ell}p^{|\mathcal{A}_i|}-C(\sum_{i=1}^{\ell}p^{|\mathcal{A}_i|-1})}{p-1}\\
=&\frac{\sum_{r=1}^{s}(\sum_{k=1}^{\ell_{r}}(-1)^{k-1}\sum_{1\leq i_1<\dots<i_k\leq \ell_{r}}p^{|\bigcap_{j=1}^{k}\mathcal{B}_{i_j}^{(r)}|}-1)}{p-1},
\end{align*}

i.e.,
\begin{equation}\label{eq:01}
C\left(\sum_{i=1}^{\ell}p^{|\mathcal{A}_i|-1}\right)=
-\sum_{r=1}^{s}\left(\sum_{k=2}^{\ell_{r}}(-1)^{k-1}\sum_{1\leq i_1<\dots<i_k\leq \ell_{r}}p^{|\bigcap_{j=1}^{k}\mathcal{B}_{i_j}^{(r)}|}-1\right).
\end{equation}
For simplicity, we use LHS to denote the left hand side of equation \eqref{eq:01}, and RHS to denote the right hand side of equation \eqref{eq:01}.
Then we can rewrite \eqref{eq:01} as ${\rm LHS}={\rm RHS}$.

It can be easily seen that for each $r\in [s]$,
$$|\mathcal{D}_{r}|=\frac{\sum_{k=1}^{\ell_{r}}(-1)^{k-1}\sum_{1\leq i_1<\dots<i_k\leq \ell_{r}}p^{|\bigcap_{j=1}^{k}\mathcal{B}_{i_j}^{(r)}|}-1}{p-1}\leq \frac{\sum_{i=1}^{\ell_{r}}p^{|\mathcal{B}_{i}^{(r)}|}-\ell_{r}}{p-1},$$
 which implies that
$$\sum_{k=2}^{\ell_{r}}(-1)^{k-1}\sum_{1\leq i_1<\dots<i_k\leq \ell_{r}}p^{|\bigcap_{j=1}^{k}\mathcal{B}_{i_j}^{(r)}|}-1\leq -\ell_{r},$$
 where the equality holds if and only if $\mathcal{B}_{1}^{(r)},\mathcal{B}_{2}^{(r)}, \dots ,\mathcal{B}_{\ell_{r}}^{(r)}$ are mutually disjoint.
Hence, ${\rm RHS}\geq \sum_{r=1}^{s}\ell_{r}=\ell$.

On the other hand, we observe that ${\rm LHS}=C(\sum_{i=1}^{\ell}p^{|\mathcal{A}_i|-1})=\sum_{i=g}^{h}b_i\leq \ell$, where the equality holds if and only if $M(\mathcal{A}_1,\mathcal{A}_2, \dots ,\mathcal{A}_{\ell}) \leq p-1$.

In summary, we have $\ell\geq {\rm LHS}$, and ${\rm RHS}\geq \ell$.
Therefore, \eqref{eq:01} holds if and only if ${\rm RHS}=\ell$ and ${\rm LHS}=\ell$, if and only if $\mathcal{B}_1^{(r)},\mathcal{B}_2^{(r)}, \dots ,\mathcal{B}_{\ell_r}^{(r)}$ are mutually disjoint for every $1\leq r\leq s$ and $M(\mathcal{A}_1,\mathcal{A}_2, \dots ,\mathcal{A}_{\ell}) \leq p- 1.$

(2) For any positive integer $t$,
\begin{align*}
\sum_{j=0}^{m-1}&\left\lceil\frac{sp^{m-1}-\sum_{i=1}^{\ell}p^{|\mathcal{A}_i|-1}+t}{p^j}\right\rceil \\
&\geq \sum_{j=0}^{m-1}\left\lceil\frac{sp^{m-1}-\sum_{i=1}^{\ell}p^{|\mathcal{A}_i|-1}+1}{p^j}\right\rceil \\   &=s\frac{p^m-1}{p-1}-\frac{\sum_{i=1}^{\ell}p^{|\mathcal{A}_i|}-C(\sum_{i=1}^{\ell}p^{|\mathcal{A}_i|-1})}{p-1}+v_{p}\left(\sum_{i=1}^{\ell}p^{|\mathcal{A}_i|-1}\right)+1.
\end{align*}

Due to $\sum_{r=1}^{s}|\mathcal{D}_{r}^{c}|>\frac{\sum_{i=1}^{\ell} p^{|\mathcal{A}_i|}-C(\sum_{i=1}^{\ell} p^{|\mathcal{A}_i|-1})}{p-1}-v_p(\sum_{i=1}^{\ell}p^{|\mathcal{A}_i|-1})-1$, we have
\begin{small}
\begin{align*}
s\frac{p^m-1}{p-1}-\sum_{r=1}^{s}|\mathcal{D}_{r}^{c}|& < s\frac{p^m-1}{p-1}-\frac{\sum_{i=1}^{\ell}p^{|\mathcal{A}_i|}-C(\sum_{i=1}^{\ell}p^{|\mathcal{A}_i|-1})}{p-1}+v_{p}\left(\sum_{i=1}^{\ell}p^{|\mathcal{A}_i|-1}\right)+1  \\
&\leq \sum_{j=0}^{m-1}\left\lceil\frac{sp^{m-1}-\sum_{i=1}^{\ell}p^{|\mathcal{A}_i|-1}+t}{p^j}\right\rceil.
\end{align*}
\end{small}
According to the Griesmer bound, there is no $p$-ary $[s\frac{p^m-1}{p-1}-\sum_{r=1}^{s}|\mathcal{D}_{r}^{c}|,m,d> sp^{m-1}-\sum_{i=1}^{\ell}p^{|\mathcal{A}_i|-1}]$
linear code. Therefore, the linear code $\mathcal{C}_{(\mathcal{D}_1,\mathcal{D}_2,\dots,\mathcal{D}_s)}$ with parameters $[s\frac{p^m-1}{p-1}-\sum_{r=1}^{s}|\mathcal{D}_{r}^{c}|, m, sp^{m-1}-\sum_{i=1}^{\ell}p^{|\mathcal{A}_i|-1}]$ is distance-optimal with respect to the Griesmer bound.

\EOP

\begin{remark}
Theorems 3.1 and 3.2 in \cite{Luo2022} can be regarded as special cases of Theorems \ref{thm:construction} and \ref{thm:iff} with $s=1$, respectively.
\end{remark}

\begin{remark}
Suppose that $\mathcal{A}_1, \mathcal{A}_2,\dots, \mathcal{A}_{\ell}$ are nonempty subsets of $[m]$ satisfying Property I$_{1}$, then for any integer $t\geq 1$, the $t$-copies of $\mathcal{A}_1, \mathcal{A}_2,\dots, \mathcal{A}_{\ell}$ satisfy Property I$_{t}$. From Theorem~\ref{thm:iff}, we can derive that
the $r$-repetition of Griesmer codes constructed from Theorem 3.2 in \cite{Luo2022} are also Griesmer codes, where $1\leq r \leq\left\lfloor\frac{p-1}{M(\mathcal{A}_1,\mathcal{A}_2, \dots ,\mathcal{A}_{\ell})}\right\rfloor$.
\end{remark}

In the follows, we will give two corollaries of Theorems~\ref{thm:construction} and \ref{thm:iff} by considering the case of $s=2$, to show that it is possible to construct good linear codes with new parameters, and that sometimes the weight distribution can be easily determined.

\begin{corollary}\label{cor:4sets}
Let $p\geq 3$ be an odd prime and let $m$ be a positive integer. Suppose that $\mathcal{A}_1$, $\mathcal{A}_2$ $\mathcal{A}_3$, $\mathcal{A}_4$ are nonempty subsets of $[m]$ such that
\begin{description}
  \item[\textnormal{(i)}] $\mathcal{A}_{3} \subseteq \mathcal{A}_{1}$, $\mathcal{A}_{4} \subseteq \mathcal{A}_{2}$,
  \item[\textnormal{(ii)}] $\mathcal{A}_{1} \cap \mathcal{A}_{2}=\emptyset$,
  \item[\textnormal{(iii)}] $M(\mathcal{A}_1, \mathcal{A}_2,\mathcal{A}_3, \mathcal{A}_4)\leq p-1$, and
  \item[\textnormal{(iv)}] $p^m > p^{|\mathcal{A}_1|} + p^{|\mathcal{A}_2|}$, $p^m > p^{|\mathcal{A}_3|}+ p^{|\mathcal{A}_4|}$.
\end{description}
If $\mathcal{D}_{1} = P_{[m]} \setminus  (P_{\mathcal{A}_1}\cup P_{\mathcal{A}_2})$,  $\mathcal{D}_{2} = P_{[m]} \setminus  (P_{\mathcal{A}_3}\cup P_{\mathcal{A}_4})$,
then $\mathcal{C}_{(\mathcal{D}_{1},\mathcal{D}_{2})}$ constructed by \eqref{eq:define} is a $p$-ary $[\frac{2p^m-\sum_{i=1}^{4}p^{|\mathcal{A}_1|}+2}{p-1} ,m, 2p^{m-1}-\sum_{i=1}^{4}p^{|\mathcal{A}_i|-1}]$
Griesmer code, whose weight distribution is listed in Table~\ref{tab1}.
\end{corollary}
\pf
From (i)-(ii), we can check that $\mathcal{A}_1$, $\mathcal{A}_2$, $\mathcal{A}_3$, $\mathcal{A}_4$ satisfy Property I$_2$. Together with  (iii)-(iv), it follows from Theorem~\ref{thm:iff}
that $\mathcal{C}_{(\mathcal{D}_{1},\mathcal{D}_{2})}$ is a Griesmer code over $\bF_p$. For any $\mathbf{x} \in \bF^{m*}_{p}$ , the weight of a codeword $c_{\mathbf{x}}$ is
$$\mbox{wt}(c_{\mathbf{x}})=2p^{m-1}-\sum_{i=1}^{4}p^{|\mathcal{A}_i|-1}+
\frac{4}{p}+\frac{1}{p}\sum_{y\in\bF_{p}^{*}}\chi_{\mathbf{x}}(y(P_{\mathcal{A}_1}\cup P_{\mathcal{A}_2}))+\frac{1}{p}\sum_{y\in\bF_{p}^{*}}\chi_{\mathbf{x}}(y(P_{\mathcal{A}_3}\cup P_{\mathcal{A}_4})).$$
According to Lemma~\ref{lem:lianggejihe}, we have $\mbox{wt}(c_{\mathbf{x}}) =$
\begin{small}
\[
\left\{
\begin{array}{lcl}
2p^{m-1},& &  {\rm if }\,\, \mathbf{x}_{\mathcal{A}_1}=\mathbf{0}, \mathbf{x}_{\mathcal{A}_2}=\mathbf{0}, \\
2p^{m-1}-p^{|\mathcal{A}_{2}|-1},& &  {\rm if }\,\, \mathbf{x}_{\mathcal{A}_1}=\mathbf{0}, \mathbf{x}_{\mathcal{A}_2}\neq\mathbf{0}, \mathbf{x}_{\mathcal{A}_4}=\mathbf{0}, \\
2p^{m-1}-p^{|\mathcal{A}_{2}|-1}-p^{|\mathcal{A}_{4}|-1},& &  {\rm if }\,\,\mathbf{x}_{\mathcal{A}_1}=\mathbf{0}, \mathbf{x}_{\mathcal{A}_2}\neq\mathbf{0}, \mathbf{x}_{\mathcal{A}_4}\neq\mathbf{0}, \\
2p^{m-1}-p^{|\mathcal{A}_{1}|-1},& &  {\rm if }\,\, \mathbf{x}_{\mathcal{A}_1}\neq\mathbf{0}, \mathbf{x}_{\mathcal{A}_2}=\mathbf{0},\mathbf{x}_{\mathcal{A}_3}=\mathbf{0}, \\
2p^{m-1}-p^{|\mathcal{A}_{1}|-1}-p^{|\mathcal{A}_{3}|-1},& &  {\rm if }\,\, \mathbf{x}_{\mathcal{A}_1}\neq\mathbf{0}, \mathbf{x}_{\mathcal{A}_2}=\mathbf{0},
\mathbf{x}_{\mathcal{A}_3}\neq\mathbf{0}, \\
2p^{m-1}-p^{|\mathcal{A}_{1}|-1}-p^{|\mathcal{A}_{2}|-1},& &  {\rm if }\,\, \mathbf{x}_{\mathcal{A}_1}\neq\mathbf{0}, \mathbf{x}_{\mathcal{A}_2}\neq\mathbf{0},
\mathbf{x}_{\mathcal{A}_3}=\mathbf{0}, \mathbf{x}_{\mathcal{A}_4}=\mathbf{0}, \\
2p^{m-1}-p^{|\mathcal{A}_{1}|-1}-p^{|\mathcal{A}_{2}|-1}-p^{|\mathcal{A}_{4}|-1},& &  {\rm if }\,\, \mathbf{x}_{\mathcal{A}_1}\neq\mathbf{0}, \mathbf{x}_{\mathcal{A}_2}\neq\mathbf{0},
\mathbf{x}_{\mathcal{A}_3}=\mathbf{0}, \mathbf{x}_{\mathcal{A}_4}\neq\mathbf{0}, \\
2p^{m-1}-p^{|\mathcal{A}_{1}|-1}-p^{|\mathcal{A}_{2}|-1}-p^{|\mathcal{A}_{3}|-1},& &  {\rm if }\,\, \mathbf{x}_{\mathcal{A}_1}\neq\mathbf{0}, \mathbf{x}_{\mathcal{A}_2}\neq\mathbf{0},
\mathbf{x}_{\mathcal{A}_3}\neq\mathbf{0}, \mathbf{x}_{\mathcal{A}_4}=\mathbf{0}, \\
2p^{m-1}-p^{|\mathcal{A}_{1}|-1}-p^{|\mathcal{A}_{2}|-1}-p^{|\mathcal{A}_{3}|-1}-p^{|\mathcal{A}_{4}|-1},& &  {\rm if }\,\, \mathbf{x}_{\mathcal{A}_1}\neq\mathbf{0}, \mathbf{x}_{\mathcal{A}_2}\neq\mathbf{0},
\mathbf{x}_{\mathcal{A}_3}\neq\mathbf{0}, \mathbf{x}_{\mathcal{A}_4}\neq\mathbf{0}.
\end{array} \right.
\]
\end{small}
The multiplicity corresponding to each weight follows.
\EOP

\begin{table}[!htb]
\caption{The weight distribution of the code in Corollary~\ref{cor:4sets}}\label{tab1}%
\resizebox{\textwidth}{20mm}{
\begin{tabular}{ll}
 \toprule
 Weights & Multiplicity \\
\midrule
1 & 0  \\
$2p^{m-1}$       &           $p^{m-|\mathcal{A}_1|-|\mathcal{A}_2|}-1$ \\

$2p^{m-1}-p^{|\mathcal{A}_{2}|-1}$  &
$p^{m-|\mathcal{A}_1|-|\mathcal{A}_4|}-p^{m-|\mathcal{A}_1|-|\mathcal{A}_2|}$  \\

$2p^{m-1}-p^{|\mathcal{A}_{2}|-1}-p^{|\mathcal{A}_{4}|-1}$ & $p^{m-|\mathcal{A}_1|}-p^{m-|\mathcal{A}_1|-|\mathcal{A}_4|}$  \\

$2p^{m-1}-p^{|\mathcal{A}_{1}|-1}$ & $p^{m-|\mathcal{A}_2|-|\mathcal{A}_3|}-p^{m-|\mathcal{A}_1|-|\mathcal{A}_2|}$  \\

$2p^{m-1}-p^{|\mathcal{A}_{1}|-1}-p^{|\mathcal{A}_{3}|-1}$ & $p^{m-|\mathcal{A}_2|}-p^{m-|\mathcal{A}_2|-|\mathcal{A}_3|}$\\

$2p^{m-1}-p^{|\mathcal{A}_{1}|-1}-p^{|\mathcal{A}_{2}|-1}$ &
$p^{m-|\mathcal{A}_1|-|\mathcal{A}_2|}(p^{|\mathcal{A}_1|-|\mathcal{A}_3|}-1)(p^{|\mathcal{A}_2|-|\mathcal{A}_4|}-1)$  \\

$2p^{m-1}-p^{|\mathcal{A}_{1}|-1}-p^{|\mathcal{A}_{2}|-1}-p^{|\mathcal{A}_{4}|-1}$ &
$p^{m-|\mathcal{A}_1|-|\mathcal{A}_2|}(p^{|\mathcal{A}_1|-|\mathcal{A}_3|}-1)(p^{|\mathcal{A}_2|}-p^{|\mathcal{A}_2|-|\mathcal{A}_4|})$    \\

$2p^{m-1}-p^{|\mathcal{A}_{1}|-1}-p^{|\mathcal{A}_{2}|-1}-p^{|\mathcal{A}_{3}|-1}$ &
$p^{m-|\mathcal{A}_1|-|\mathcal{A}_2|}(p^{|\mathcal{A}_2|-|\mathcal{A}_4|}-1)(p^{|\mathcal{A}_1|}-p^{|\mathcal{A}_1|-|\mathcal{A}_3|})$      \\

$2p^{m-1}-p^{|\mathcal{A}_{1}|-1}-p^{|\mathcal{A}_{2}|-1}-p^{|\mathcal{A}_{3}|-1}-p^{|\mathcal{A}_{4}|-1}$ &
$p^{m-|\mathcal{A}_1|-|\mathcal{A}_2|}(p^{|\mathcal{A}_1|}-p^{|\mathcal{A}_1|-|\mathcal{A}_3|})(p^{|\mathcal{A}_2|}-p^{|\mathcal{A}_2|-|\mathcal{A}_4|})$\\
 \bottomrule
\end{tabular}
}
\end{table}

Below we give an example to illustrate Corollary \ref{cor:4sets}.
\begin{example}
Let $p = m = 3$, $\mathcal{A}_1 = \{1, 2\}$, $\mathcal{A}_2 = \{3\}$, $\mathcal{A}_3 = \{1\}$, $\mathcal{A}_4 = \emptyset$. Let $\mathcal{D}_{1} = P_{[m]} \setminus  (P_{\mathcal{A}_1}\cup P_{\mathcal{A}_2})$,  $\mathcal{D}_{2} = P_{[m]} \setminus  (P_{\mathcal{A}_3}\cup P_{\mathcal{A}_4})$, $\mathcal{C}_{(\mathcal{D}_{1},\mathcal{D}_{2})}$ be the $3$-ary code constructed by \eqref{eq:define},  from Corollary~\ref{cor:4sets},  we know $\mathcal{C}_{(\mathcal{D}_{1},\mathcal{D}_{2})}$ is a Griesmer code  $[20,3,13]_3$ with weight enumerator $1+12x^{13}+10x^{14}+2x^{15}+2x^{17}$.
%which is identical to the best known linear code from the Magma BKLC $(\bF_3, 20, 3)$.
\end{example}

\begin{corollary}\label{cor:intersect}
Let $p$ be an odd prime and $m$ a positive integer. Suppose that $\mathcal{A}_1$, $\mathcal{A}_2$,  $\mathcal{A}_3$, $\mathcal{A}_4$ are nonempty subsets of $[m]$ such that
\begin{description}
  \item[\textnormal{(i)}] $\mathcal{A}_{3} \subseteq \mathcal{A}_{1}$, $\mathcal{A}_{4} \subseteq \mathcal{A}_{2}$,
  \item[\textnormal{(ii)}] $\mathcal{A}_{1} \cap \mathcal{A}_{2}\neq\emptyset$,
  \item[\textnormal{(iii)}] $\mathcal{A}_{3}\not\subseteq \mathcal{A}_{1}\cap\mathcal{A}_{2}$, $\mathcal{A}_{4}\not\subseteq \mathcal{A}_{1}\cap\mathcal{A}_{2}$, and
  \item[\textnormal{(iv)}] $p^m > p^{|\mathcal{A}_1|} + p^{|\mathcal{A}_2|}$.
\end{description}
If $\mathcal{D}_{1} = P_{[m]} \setminus  (P_{\mathcal{A}_1}\cup P_{\mathcal{A}_2})$,  $\mathcal{D}_{2} = P_{[m]} \setminus  (P_{\mathcal{A}_3}\cup P_{\mathcal{A}_4})$,
then $\mathcal{C}_{(\mathcal{D}_{1},\mathcal{D}_{2})}$ constructed by \eqref{eq:define} is a $p$-ary linear code with parameters
$$[\frac{2p^m-\sum_{i=1}^{4}p^{|\mathcal{A}_i|}+p^{|\mathcal{A}_1\cap \mathcal{A}_2|}+p^{|\mathcal{A}_3\cap \mathcal{A}_4|}}{p-1} ,m, 2p^{m-1} -\sum_{i=1}^{4} p^{|\mathcal{A}_i|-1}].$$
Furthermore, without loss of generality, let $|\mathcal{A}_3|=\min\{|\mathcal{A}_i|: i=1,2,3,4\}$,  then $\mathcal{C}_{(\mathcal{D}_{1},\mathcal{D}_{2})}$ is distance-optimal with respect to the Griesmer bound if any one of the following conditions is satisfied.
\begin{itemize}
  \item[\textnormal{(1)}] $|\mathcal{A}_3| >\frac{p^{|\mathcal{A}_1\cap \mathcal{A}_2|}+p^{|\mathcal{A}_3\cap \mathcal{A}_4|}-2}{p-1}$, if $M(\mathcal{A}_1,\mathcal{A}_2,\mathcal{A}_3,\mathcal{A}_4)\leq p-1$.
  \item[\textnormal{(2)}] $|\mathcal{A}_3| >\frac{p^{|\mathcal{A}_1\cap \mathcal{A}_2|}+p^{|\mathcal{A}_3\cap \mathcal{A}_4|}}{p-1}$, if $p=3$,  $|\mathcal{A}_1|=|\mathcal{A}_2|=|\mathcal{A}_4|>|\mathcal{A}_3|$.
  \item[\textnormal{(3)}] $|\mathcal{A}_3| >\frac{p^{|\mathcal{A}_1\cap \mathcal{A}_2|}+p^{|\mathcal{A}_3\cap \mathcal{A}_4|}}{p-1}-1$, if $p=3$, $|\mathcal{A}_1|>|\mathcal{A}_2|=|\mathcal{A}_3|=|\mathcal{A}_4|$.
  \item[\textnormal{(4)}] $|\mathcal{A}_3|>\frac{p^{|\mathcal{A}_1\cap \mathcal{A}_2|}+p^{|\mathcal{A}_3\cap\mathcal{A}_4|}}{p-1}-1$, if $p=3$, $|\mathcal{A}_1|=|\mathcal{A}_2|=|\mathcal{A}_3|=|\mathcal{A}_4|$.
\end{itemize}

\end{corollary}

\pf
From conditions (i)-(iv) and Theorem~\ref{thm:construction}, the parameters of $\mathcal{C}_{(\mathcal{D}_{1},\mathcal{D}_{2})}$ follows.\\
(1) When $M(\mathcal{A}_1,\mathcal{A}_2,\mathcal{A}_3,\mathcal{A}_4)\leq p-1$, it is easy to see that
$C(\sum_{i=1}^{4}p^{|\mathcal{A}_i|-1})=4$, $v_{p}(\sum_{i=1}^{4}p^{|\mathcal{A}_i|-1})=|\mathcal{A}_3|-1$.
Due to $|\mathcal{A}_3| >\frac{p^{|\mathcal{A}_1\cap \mathcal{A}_2|}+p^{|\mathcal{A}_3\cap \mathcal{A}_4|}-2}{p-1}$, we obtain that
\begin{align*}
|\mathcal{D}_{1}^{c}|+|\mathcal{D}_{2}^{c}|&=\frac{\sum_{i=1}^{4}p^{|\mathcal{A}_i|}-p^{|\mathcal{A}_1\cap \mathcal{A}_2|}-p^{|\mathcal{A}_3\cap \mathcal{A}_4|}-2}{p-1}\\
&>\frac{\sum_{i=1}^{4}p^{|\mathcal{A}_i|}-4}{p-1}-|\mathcal{A}_3|\\
&= \frac{\sum_{i=1}^{4} p^{|\mathcal{A}_i|}-C(\sum_{i=1}^{4} p^{|\mathcal{A}_i|-1})}{p-1}-v_{p}\left(\sum_{i=1}^{4}p^{|\mathcal{A}_i|-1}\right)-1.
\end{align*}
According to Theorem~\ref{thm:iff}, $\mathcal{C}_{(\mathcal{D}_{1},\mathcal{D}_{2})}$ is distance-optimal with respect to the Griesmer bound.\\
(2-4) Since $M(\mathcal{A}_1,\mathcal{A}_2,\mathcal{A}_3,\mathcal{A}_4)\leq 4$, $M(\mathcal{A}_1,\mathcal{A}_2,\mathcal{A}_3,\mathcal{A}_4)> p-1$ if and only if $p=3$.
\begin{align*}
M(\mathcal{A}_1,\mathcal{A}_2,\mathcal{A}_3,\mathcal{A}_4)=3 &\Leftrightarrow
\left\{
\begin{array}{lcl}
|\mathcal{A}_1|=|\mathcal{A}_2|=|\mathcal{A}_4|>|\mathcal{A}_3|\\
|\mathcal{A}_1|>|\mathcal{A}_2|=|\mathcal{A}_3|=|\mathcal{A}_4|
 \end{array} \right.\\
&\Leftrightarrow
\left\{
 \begin{array}{lcl}
C(\sum_{i=1}^{4}p^{|\mathcal{A}_i|-1})=2, v_{p}(\sum_{i=1}^{4}p^{|\mathcal{A}_i|-1})=|\mathcal{A}_3|-1.\\
C(\sum_{i=1}^{4}p^{|\mathcal{A}_i|-1})=2, v_{p}(\sum_{i=1}^{4}p^{|\mathcal{A}_i|-1})=|\mathcal{A}_3|.
\end{array} \right.
\end{align*}
\begin{align*}
M(\mathcal{A}_
1,\mathcal{A}_2,\mathcal{A}_3,\mathcal{A}_4)=4 &\Leftrightarrow
|\mathcal{A}_1|=|\mathcal{A}_2|=|\mathcal{A}_4|=|\mathcal{A}_3|\\
&\Leftrightarrow
C\left(\sum_{i=1}^{4}p^{|\mathcal{A}_i|-1}\right)=2, v_{p}\left(\sum_{i=1}^{4}p^{|\mathcal{A}_i|-1}\right)=|\mathcal{A}_3|.
\end{align*}

These three circumstances corresponding to (2-4) respectively, the remaining proofs of (2-4) can be done similarly to (1).
\EOP

In general, it is complicated to calculate the weight distribution of code constructed in Corollary~\ref{cor:intersect}, as we need to consider the following $15$ cases:
\[\left\{
\begin{array}{l}
\mathbf{x}_{\mathcal{A}_1}=\mathbf{0}, \mathbf{x}_{\mathcal{A}_2}=\mathbf{0}, \\
\mathbf{x}_{\mathcal{A}_1}=\mathbf{0}, \mathbf{x}_{\mathcal{A}_2}\neq\mathbf{0},\left\{
  \begin{array}{l}
  \mathbf{x}_{\mathcal{A}_4}=\mathbf{0},\\
  \mathbf{x}_{\mathcal{A}_4}\neq\mathbf{0},
  \end{array} \right. \\
\mathbf{x}_{\mathcal{A}_1}\neq\mathbf{0}, \mathbf{x}_{\mathcal{A}_2}=\mathbf{0},\left\{
  \begin{array}{l}
  \mathbf{x}_{\mathcal{A}_3}=\mathbf{0},\\
  \mathbf{x}_{\mathcal{A}_3}\neq\mathbf{0},
  \end{array} \right. \\
\mathbf{x}_{\mathcal{A}_1}\neq\mathbf{0}, \mathbf{x}_{\mathcal{A}_2}\neq\mathbf{0},
  \left\{
  \begin{array}{l}
  \mathbf{x}_{\mathcal{A}_3}=\mathbf{0},\mathbf{x}_{\mathcal{A}_4}=\mathbf{0},
     \left\{
     \begin{array}{l}
     \mathbf{x}_{\mathcal{A}_1\cap \mathcal{A}_2}=\mathbf{0},\\
     \mathbf{x}_{\mathcal{A}_1\cap \mathcal{A}_2}\neq\mathbf{0},
     \end{array} \right.\\
  \mathbf{x}_{\mathcal{A}_3}=\mathbf{0},\mathbf{x}_{\mathcal{A}_4}\neq\mathbf{0},
     \left\{
     \begin{array}{l}
     \mathbf{x}_{\mathcal{A}_1\cap \mathcal{A}_2}=\mathbf{0},\\
     \mathbf{x}_{\mathcal{A}_1\cap \mathcal{A}_2}\neq\mathbf{0},
     \end{array} \right. \\
  \mathbf{x}_{\mathcal{A}_3}\neq\mathbf{0}, \mathbf{x}_{\mathcal{A}_4}=\mathbf{0},
     \left\{
     \begin{array}{l}
     \mathbf{x}_{\mathcal{A}_1\cap \mathcal{A}_2}=\mathbf{0},\\
     \mathbf{x}_{\mathcal{A}_1\cap \mathcal{A}_2}\neq\mathbf{0},
     \end{array} \right.\\
  \mathbf{x}_{\mathcal{A}_3}\neq\mathbf{0}, \mathbf{x}_{\mathcal{A}_4}\neq\mathbf{0},
     \left\{\begin{array}{l}
     \mathbf{x}_{\mathcal{A}_3\cap \mathcal{A}_4}=\mathbf{0},
     \left\{
         \begin{array}{l}
         \mathbf{x}_{\mathcal{A}_1\cap \mathcal{A}_2}=\mathbf{0},\\
         \mathbf{x}_{\mathcal{A}_1\cap \mathcal{A}_2}\neq\mathbf{0},
         \end{array} \right. \\
     \mathbf{x}_{\mathcal{A}_3\cap \mathcal{A}_4}\neq\mathbf{0},
         \left\{
         \begin{array}{l}
         \mathbf{x}_{\mathcal{A}_1\cap \mathcal{A}_2}=\mathbf{0},\\
         \mathbf{x}_{\mathcal{A}_1\cap \mathcal{A}_2}\neq\mathbf{0}.
         \end{array} \right.
    \end{array} \right.
\end{array} \right.
 \end{array} \right.  \]

Correspondingly, $\mbox{wt}(c_{\mathbf{x}})$ is equal to
\[\left\{
\begin{array}{l}
2p^{m-1}, \\
2p^{m-1}-p^{|\mathcal{A}_2|-1},\\
2p^{m-1}-p^{|\mathcal{A}_2|-1}-p^{|\mathcal{A}_4|-1},\\
2p^{m-1}-p^{|\mathcal{A}_1|-1},\\
2p^{m-1}-p^{|\mathcal{A}_1|-1}-p^{|\mathcal{A}_3|-1},\\
2p^{m-1}-p^{|\mathcal{A}_1|-1}-p^{|\mathcal{A}_2|-1},\\
2p^{m-1}-p^{|\mathcal{A}_1|-1}-p^{|\mathcal{A}_2|-1}+p^{|\mathcal{A}_1\cap\mathcal{A}_2|-1},\\
2p^{m-1}-p^{|\mathcal{A}_1|-1}-p^{|\mathcal{A}_2|-1}-p^{|\mathcal{A}_4|-1},\\
2p^{m-1}-p^{|\mathcal{A}_1|-1}-p^{|\mathcal{A}_2|-1}-p^{|\mathcal{A}_4|-1}+p^{|\mathcal{A}_1\cap\mathcal{A}_2|-1},\\
2p^{m-1}-p^{|\mathcal{A}_1|-1}-p^{|\mathcal{A}_2|-1}-p^{|\mathcal{A}_3|-1},\\
2p^{m-1}-p^{|\mathcal{A}_1|-1}-p^{|\mathcal{A}_2|-1}-p^{|\mathcal{A}_3|-1}+p^{|\mathcal{A}_1\cap\mathcal{A}_2|},\\
2p^{m-1}-p^{|\mathcal{A}_1|-1}-p^{|\mathcal{A}_2|-1}-p^{|\mathcal{A}_3|-1}-p^{|\mathcal{A}_4|-1},\\
2p^{m-1}-p^{|\mathcal{A}_1|-1}-p^{|\mathcal{A}_2|-1}-p^{|\mathcal{A}_3|-1}-p^{|\mathcal{A}_4|-1}+p^{|\mathcal{A}_1\cap\mathcal{A}_2|-1},\\
2p^{m-1}-p^{|\mathcal{A}_1|-1}-p^{|\mathcal{A}_2|-1}-p^{|\mathcal{A}_3|-1}-p^{|\mathcal{A}_4|-1}+p^{|\mathcal{A}_3\cap\mathcal{A}_4|-1},\\
2p^{m-1}-p^{|\mathcal{A}_1|-1}-p^{|\mathcal{A}_2|-1}-p^{|\mathcal{A}_3|-1}-p^{|\mathcal{A}_4|-1}+p^{|\mathcal{A}_3\cap\mathcal{A}_4|-1}+p^{|\mathcal{A}_1\cap\mathcal{A}_2|-1}.
\end{array} \right.  \]

We will not give the explicit expression of the weight distribution here, but we will illustrate the computation with an example.

\begin{example}
Let $p = 3$, $m=4$ and $\mathcal{A}_1 = \{1, 2, 3\}$, $\mathcal{A}_2 = \{3, 4\}$, $\mathcal{A}_3 = \{1, 2\}$, $\mathcal{A}_4 = \{3, 4\}$. Let $\mathcal{D}_{1} = P_{[m]} \setminus  (P_{\mathcal{A}_1}\cup P_{\mathcal{A}_2})$,  $\mathcal{D}_{2} = P_{[m]} \setminus  (P_{\mathcal{A}_3}\cup P_{\mathcal{A}_4})$, $\mathcal{C}_{(\mathcal{D}_{1},\mathcal{D}_{2})}$ be the $3$-ary code constructed by \eqref{eq:define}. From Corollary~\ref{cor:intersect}, we know that the parameters of $\mathcal{C}_{(\mathcal{D}_{1},\mathcal{D}_{2})}$ are $[56,4,36]_3$. Referring to the above analysis, for $\mathbf{x}\in\bF_{3}^{4*}$, we have
\[\mbox{wt}(c_{\mathbf{x}})=\left\{
\begin{array}{ll}
2\cdot3^{3}-3-3,&\mathbf{x}_{\mathcal{A}_1}=\mathbf{0}, \mathbf{x}_{\mathcal{A}_2}\neq\mathbf{0}, \\
2\cdot3^{3}-3^{2},&\mathbf{x}_{\mathcal{A}_1}\neq\mathbf{0}, \mathbf{x}_{\mathcal{A}_2}=\mathbf{0}, \mathbf{x}_{\mathcal{A}_3}=\mathbf{0}, \\
2\cdot3^{3}-3^{2}-3,&\mathbf{x}_{\mathcal{A}_1}\neq\mathbf{0}, \mathbf{x}_{\mathcal{A}_2}=\mathbf{0}, \mathbf{x}_{\mathcal{A}_3}\neq\mathbf{0},\\
2\cdot3^{3}-3^{2}-3-3,&\mathbf{x}_{\mathcal{A}_1}\neq\mathbf{0}, \mathbf{x}_{\mathcal{A}_2}\neq\mathbf{0}, \mathbf{x}_{\mathcal{A}_3}=\mathbf{0},\mathbf{x}_{\mathcal{A}_1\cap \mathcal{A}_2}=\mathbf{0},\\
2\cdot3^{3}-3^{2}-3-3+3,&\mathbf{x}_{\mathcal{A}_1}\neq\mathbf{0}, \mathbf{x}_{\mathcal{A}_2}\neq\mathbf{0}, \mathbf{x}_{\mathcal{A}_3}=\mathbf{0},\mathbf{x}_{\mathcal{A}_1\cap \mathcal{A}_2}\neq\mathbf{0},\\
2\cdot3^{3}-3^{2}-3-3-3,&\mathbf{x}_{\mathcal{A}_1}\neq\mathbf{0}, \mathbf{x}_{\mathcal{A}_2}\neq\mathbf{0}, \mathbf{x}_{\mathcal{A}_3}\neq\mathbf{0},\mathbf{x}_{\mathcal{A}_1\cap \mathcal{A}_2}=\mathbf{0},\\
2\cdot3^{3}-3^{2}-3-3-3+3,&\mathbf{x}_{\mathcal{A}_1}\neq\mathbf{0}, \mathbf{x}_{\mathcal{A}_2}\neq\mathbf{0}, \mathbf{x}_{\mathcal{A}_3}\neq\mathbf{0},\mathbf{x}_{\mathcal{A}_1\cap \mathcal{A}_2}\neq\mathbf{0}.
\end{array} \right.  \]
So $\mathcal{C}_{(\mathcal{D}_{1},\mathcal{D}_{2})}$ here is a $5$-weight code with nonzero weights $\{36,39,42,45,48\}$.
According to the best known linear code from the Magma BKLC $(\bF_3, 56, 4)$ with weight enumerator $1+76x^{36}+4x^{45}$, it can be seen that $\mathcal{C}_{(\mathcal{D}_{1},\mathcal{D}_{2})}$  is a new distance-optimal linear code.
\end{example}

\section{Alphabet-optimal $(r,\delta)$-LRCs}\label{sec:locality}

In this section, we will revisit the locality of codes constructed in \cite{Luo2022}, i.e., the codes in Theorem~\ref{thm:construction} with $s=1$. It turns out that the constructed codes in \cite{Luo2022}, which are alphabet-optimal $2$-LRCs, can also be characterized as alphabet-optimal $(2,p-1)$-LRCs. Furthermore, we will investigate the conditions under which the codes constructed from \eqref{eq:framework} possess $(2, p)$ and $(2, p-2)$ localities. Some new alphabet-optimal $(r,\delta)$-LRCs are also provided.

Recall that
\begin{align*}
P_{[m]} = &\{(1, a_2, a_3, \dots , a_m) : a_2, \dots , a_m \in \bF_p\} \cup \\
&\{(0, 1, a_3, \dots , a_m) : a_3, \dots , a_m \in \bF_p\}
\cup \dots\cup \{(0, 0, \dots , 1)\}.
\end{align*}
Let $\widetilde{P}_{[m]}=\{(1, a_2, \dots , a_m) : a_2, \dots , a_m \in \bF_p^{*}\}$.
When $p^m > \sum_{i=1}^{\ell} p^{|\mathcal{A}_i|}$, which
implies that $\mathcal{A}_1,\mathcal{A}_2, \dots ,\mathcal{A}_{\ell}$ are proper subsets of $[m]$, then $\widetilde{P}_{[m]}$ is a subset of $\mathcal{D}= P_{[m]} \setminus \bigcup_{i=1}^{\ell}P_{\mathcal{A}_{i}}$, as $\widetilde{P}_{[m]}\cap P_{\mathcal{A}_{i}}=\emptyset$ for any $1\leq i\leq \ell$.

Next, we will give a formal definition of $(r,\delta)$-LRCs from the aspect of generator matrix.
\begin{definition}\label{def:rdelta}
Let $\mathcal{C}$ be a $p$-ary linear code with generator matrix
$G = [\mathbf{g}_1, \dots, \mathbf{g}_n]$.
The $i$-th coordinate, $1 \leq i \leq n$, of $\mathcal{C}$ is said to have $(r,\delta)$-locality if there exists a subset $\mathcal{I}\subset\{\mathbf{g}_1, \dots, \mathbf{g}_n\}$ containing $\mathbf{g}_i$ such that
\begin{itemize}
  \item[(1)] $|\mathcal{I}| \leq r+\delta-1$,
  \item[(2)] any $\delta-1$ vectors in $\mathcal{I}$ are  linear combinations of the remaining vectors in $\mathcal{I}$.
\end{itemize}
If all the coordinates of $\mathcal{C}$ have $(r,\delta)$-locality, then $\mathcal{C}$ is called an $(r,\delta)$-locally repairable code, or in short, $(r,\delta)$-LRC.
\end{definition}

As previously mentioned, $P_{[m]}$ can be considered as a subset of $L_{[m]}$. Therefore, we can refer to the elements of $P_{[m]}$ as vectors without any ambiguity. In $P_{[m]}$, the first nonzero coordinate of any vector is $1$, which means that the linear combination of vectors in $P_{[m]}$ may not necessarily belong to $P_{[m]}$. For the sake of convenience in later discussions, for any $\mathbf{f}\in L_{[m]}$, we denote by $[\mathbf{f}]$ the vector equivalent to $\mathbf{f}$ whose first nonzero coordinate is $1$.

\subsection{$(2,p-2)$-LRCs }\label{subsec:p-2}

In this subsection, we will explore the $(2,p-2)$-locality of $p$-ary codes constructed by \eqref{eq:framework} via analysing the linear dependence among vectors in the defining sets.
\begin{lemma}\label{lem:vectorexist}
Let $p\geq 5$ be an odd prime, $\mathbb{F}_{p}=\{0,-1,\alpha_1,\alpha_2,\dots,\alpha_{p-2}\}$.
\begin{itemize}
  \item[(1)] For any $\mathbf{g}\in \widetilde{P}_{[m]}$, there exists $\mathbf{h} \in \widetilde{P}_{[m]}$, $\mathbf{h}\neq \mathbf{g}$, such that
$$[\mathbf{h}+\alpha_{i}\mathbf{g}]\in \widetilde{P}_{[m]}$$
for all  $i \in [p-2]\setminus \{j\}$, where $j$ is arbitrary chosen from $[p-2]$.
  \item[(2)] For any $\mathbf{g}\in P_{[m]}\setminus \widetilde{P}_{[m]}$, there exists $\mathbf{h}\in \widetilde{P}_{[m]}$ such that
$$[\mathbf{h}+\alpha_{i}\mathbf{g}]\in \widetilde{P}_{[m]}$$
for all  $i \in [p-2]$.
\end{itemize}

\end{lemma}
\pf
(1) For $\mathbf{g}\in \widetilde{P}_{[m]}$, we can write $\mathbf{g} = (1, g_2,\dots, g_m)$, where $g_{i}\neq 0$ for every $i\in \{2,\dots,m\}$.
For any $j\in [p-2]$, if we choose $\mathbf{h}=(1,-\alpha_{j}g_2,\dots,-\alpha_{j}g_{m})\in \widetilde{P}_{[m]}$, then
$$[\mathbf{h}+\alpha_{i}\mathbf{g}]=\frac{1}{1+\alpha_i}(1+\alpha_i, (\alpha_i-\alpha_j)g_2,\dots,(\alpha_i-\alpha_j)g_m)\in \widetilde{P}_{[m]}$$
for all  $i \in [p-2]\setminus \{j\}$.

(2) For $\mathbf{g}\in P_{[m]}\setminus \widetilde{P}_{[m]}$, denote $i$ the position of first nonzero component of $\mathbf{g}$, where $1\leq i\leq m$. Then we can express $\mathbf{g}$ as follows
$$\mathbf{g} = (\overbrace{0,\dots,0}^{i-1},1, g_{i+1},\dots, g_m).$$
Since $\mathbf{g}\notin \widetilde{P}_{[m]}$, there is a subset $Z\subseteq \{i+1,i+2,\dots,m\}$ such that $g_{j}=0$ for $j\in Z$,  $g_{j}\neq 0$ for $j\in \{i+1,i+2,\dots,m\}\setminus Z$, where $i-1+|Z|\geq 1$.

Let $\mathbf{h}$ be an arbitrary element in
$$\{(\overbrace{1,\dots,1}^{i}, h_{i+1},\dots, h_m): h_{r}\neq 0 {~\rm if~} r\in Z, h_{r}= g_{r}  {~\rm if~} r\in \{i+1,i+2,\dots,m\}\setminus Z\}.$$
It is easy to check that $\mathbf{h}\in \widetilde{P}_{[m]}$ and
$$[\mathbf{h}+\alpha_{i}\mathbf{g}]\in \widetilde{P}_{[m]}$$
for all  $i \in [p-2]$.
\EOP

\bigskip
Below we give an example to illustrate Lemma~\ref{lem:vectorexist}.

\begin{example}
Let $p=5$, $m=4$, $\mathbb{F}_{p}=\{0,-1,\alpha_1,\alpha_2,\dots,\alpha_{p-2}\}=\{0,-1,1,2,3\}$.
\begin{itemize}
  \item[(1)] For $\mathbf{g} = (1, 2, 3, 4)$, let $j=3$, from Lemma~\ref{lem:vectorexist}, we can choose $\mathbf{h}=(1,4,1,3)$, and one can check that $[\mathbf{h}+\mathbf{g}]\in \widetilde{P}_{[4]}$, $[\mathbf{h}+2\mathbf{g}]\in \widetilde{P}_{[4]}$, $[\mathbf{h}+3\mathbf{g}]\notin \widetilde{P}_{[4]}$.
  \item[(2)] For $\mathbf{g} = (1, 0, 3, 4)$, from Lemma~\ref{lem:vectorexist}, we can choose $\mathbf{h}=(1,1,3,4)$, and one can check that $[\mathbf{h}+\mathbf{g}]\in \widetilde{P}_{[4]}$, $[\mathbf{h}+2\mathbf{g}]\in \widetilde{P}_{[4]}$, $[\mathbf{h}+3\mathbf{g}]\in \widetilde{P}_{[4]}$.
  \item[(3)]  For $\mathbf{g} = (0, 1, 0, 4)$, from Lemma~\ref{lem:vectorexist}, we can choose $\mathbf{h}=(1,1,3,4)$, and one can check that $[\mathbf{h}+\mathbf{g}]\in \widetilde{P}_{[4]}$, $[\mathbf{h}+2\mathbf{g}]\in \widetilde{P}_{[4]}$, $[\mathbf{h}+3\mathbf{g}]\in \widetilde{P}_{[4]}$.
\end{itemize}

\end{example}

By combining Definition~\ref{def:rdelta} and Lemma~\ref{lem:vectorexist}, we can establish a criterion for determining the $(2,p-2)$-locality of codes constructed from \eqref{eq:framework}.

\begin{theorem}\label{thm:locality}
Let $p\geq 5$ be an odd prime, $\ell\geq 1$ and $m\geq 2$ be integers. Suppose that $\mathcal{A}_1, \mathcal{A}_2, \dots, \mathcal{A}_{\ell}$ are proper subsets of $[m]$ such that they do not contain each other. Let $\mathcal{D}= P_{[m]} \setminus  (\bigcup_{i=1}^{\ell} P_{\mathcal{A}_i})$, then $\mathcal{C}_{\mathcal{D}}$ constructed by \eqref{eq:framework} is a $(2,p-2)$-LRC.
\end{theorem}
\pf Since $\mathcal{A}_1, \mathcal{A}_2, \dots, \mathcal{A}_{\ell}$ are proper subsets of $[m]$,   $\widetilde{P}_{[m]}\subset \mathcal{D}$.
From Lemma~\ref{lem:vectorexist}, for each $\mathbf{g}\in \widetilde{P}_{[m]}$, let $j=p-2$, then there exists $\mathbf{h} \in \widetilde{P}_{[m]}$, $\mathbf{h}\neq \mathbf{g}$, and a set $\mathcal{I}=\{\mathbf{g},\mathbf{h},[\mathbf{h}+\alpha_{1}\mathbf{g}],\dots, [\mathbf{h}+\alpha_{p-3}\mathbf{g}]\} \subset \widetilde{P}_{[m]}\subset \mathcal{D}$ of size $p-1$ such that any 2 vectors in $\mathcal{I}$ can be linearly combined to get the remaining  vectors in $\mathcal{I}$. From Definition~\ref{def:rdelta}, coordinates occupied by $\widetilde{P}_{[m]}$ have $(2,p-2)$-locality.
Similarly, we can prove that coordinates occupied by $\mathcal{D}\setminus \widetilde{P}_{[m]}$ have $(2,p-1)$-locality. Overall, $\mathcal{C}_{\mathcal{D}}$ is a $(2,p-2)$-LRC.

\EOP

\begin{remark}
The part (1) of Theorem 4.2 in \cite{Luo2022} can be obtained by Theorem~\ref{thm:locality} .
\end{remark}

\subsection{$(2,p-1)$-LRCs }
In this subsection, after adding an extra requirement to the defining sets and utilizing the results in Subsection~\ref{subsec:p-2}, we will determine the $(2,p-1)$-locality of $p$-ary codes constructed by \eqref{eq:framework}.

\begin{theorem}\label{thm:locality02}
Let $p\geq 3$ be an odd prime, $\ell\geq 1$ and $m\geq 2$ be integers.
Suppose that $\mathcal{A}_1, \mathcal{A}_2, \dots, \mathcal{A}_{\ell}$ are proper subsets of $[m]$ such that they do not contain each other, let $\mathcal{D}= P_{[m]} \setminus  (\bigcup_{i=1}^{\ell} P_{\mathcal{A}_i})$. If there exists a subset $\mathcal{A}^{*}\subset [m]$ with size $m-1$ such that $\mathcal{A}^{*}\neq \mathcal{A}_i$ for all $i\in [\ell]$,
then $\mathcal{C}_{\mathcal{D}}$ constructed by \eqref{eq:framework} is a $(2,p-1)$-LRC.
\end{theorem}
\pf
Let $\mathbb{F}_{p}=\{0,-1,\alpha_1,\alpha_2,\dots,\alpha_{p-2}\}$. According to Lemma~\ref{lem:vectorexist} and Theorem~\ref{thm:locality}, if we can show that for any $\mathbf{g}\in \widetilde{P}_{[m]}\subset \mathcal{D}$, there exists $\mathbf{h}\in  \mathcal{D}$ such that
$$\left|\{\mathbf{g},\mathbf{h},[\mathbf{h}-\mathbf{g}], [\mathbf{h}+\alpha_{1}\mathbf{g}], \dots,  [\mathbf{h}+\alpha_{p-2}\mathbf{g}]\}\cap \mathcal{D}\right|\geq p,$$
the proof is done.

For simplicity, define $[\mathbf{g},\mathbf{h}]:=\{\mathbf{g},\mathbf{h},[\mathbf{h}-\mathbf{g}], [\mathbf{h}+\alpha_{1}\mathbf{g}], \dots,  [\mathbf{h}+\alpha_{p-2}\mathbf{g}]\}$. We can see that in $[\mathbf{g},\mathbf{h}]$, any $2$ vectors can be linearly combined to obtain all the remaining vectors.

Next, we prove the theorem by considering the following two cases.

\textbf{Case (i)}: Suppose $\mathcal{A}^{*}=\{2,3,\dots,m\}$.

For any $\mathbf{g}=(1,g_2,\dots,g_m)\in \widetilde{P}_{[m]}$, where $g_{i}\neq 0$ for all $2\leq i\leq m$, let $\mathbf{h}=(h_1,h_2,\dots,h_m)=g_{2}^{-1}(0,g_2,\dots,g_m)$. For each $i\in [\ell],$ since $\mathcal{A}^{*}\setminus \mathcal{A}_i\neq \emptyset$, there exists $t_{i}\in \mathcal{A}^{*}$ such that $t_{i}\notin  \mathcal{A}_{i}$. As $h_{t_{i}}\neq 0$, we have $\mathbf{h}\notin P_{\mathcal{A}_{i}}$ for any $i\in [\ell]$.  Hence $\mathbf{h}\notin \bigcup_{j=1}^{\ell}P_{\mathcal{A}_{j}}$, i.e., $\mathbf{h}\in \mathcal{D}$.
It is easy to verify that $[\mathbf{h}+(-g_2^{-1}\mathbf{g})]=(1,0,\dots,0)$ is the only possible element in $[\mathbf{g},\mathbf{h}]$ which does not belong to $\mathcal{D}$, and all the remaining elements in $[\mathbf{g},\mathbf{h}]$ belong to $\widetilde{P}_{[m]}$. So $$\left|[\mathbf{g},\mathbf{h}]\cap \mathcal{D}\right|\geq p.$$

\textbf{Case (ii)}: Suppose $\mathcal{A}^{*}=[m]\setminus \{j\}$ for some $j\in\{2, \dots, m\}$.

For any $\mathbf{g}=(1,g_2,\dots,g_m)\in \widetilde{P}_{[m]}$, where $g_{i}\neq 0$ for all $2\leq i\leq m$, let $\mathbf{h}=(1,h_2, h_3,\dots,h_m)$ with $h_i=g_i$ for $i\in \mathcal{A}^{*}\setminus \{1\}$, and $h_{j}=0$.
Similarly, we can check that $\mathbf{h}\in \mathcal{D}$,
$$[\mathbf{h}-\mathbf{g}] = (\overbrace{0,\dots,0}^{j-1},1, 0,\dots, 0)$$
is the only possible element in $[\mathbf{g},\mathbf{h}]$ which does not belong to $\mathcal{D}$, and all the remaining  elements in $[\mathbf{g},\mathbf{h}]$ belong to $\widetilde{P}_{[m]}$. So $$\left|[\mathbf{g},\mathbf{h}]\cap \mathcal{D}\right|\geq p.$$

In summary, the proof is completed.
\EOP

\begin{remark}
The part (2) of Theorem 4.2 in \cite{Luo2022} is a special case of Theorem~\ref{thm:locality02} with $p=3$.
\end{remark}

Below we give an example to illustrate Theorem~\ref{thm:locality02}.

\begin{example}
Let $p=5$, $m=4$, $\mathbb{F}_{p}=\{0,-1,\alpha_1,\alpha_2,\alpha_{3}\}=\{0,-1,1,2,3\}$.
\begin{itemize}
  \item[(1)] For $\mathcal{A}^{*}=\{2,3,4\}$, $\mathbf{g} = (1, 2, 3, 4)$, from Theorem~\ref{thm:locality02}, we can choose $\mathbf{h}=(0,1,4,2)$, and one can check that $[\mathbf{h}+\mathbf{g}]\in \widetilde{P}_{[4]}$, $[\mathbf{h}+3\mathbf{g}]\in \widetilde{P}_{[4]}$, $[\mathbf{h}-\mathbf{g}]\in \widetilde{P}_{[4]}$.
  \item[(2)] For $\mathcal{A}^{*}=\{1,3,4\}$, $\mathbf{g} = (1, 2, 3, 4)$, from Theorem~\ref{thm:locality02}, we can choose $\mathbf{h}=(1,0,3,4)$, and one can check that $[\mathbf{h}+\mathbf{g}]\in \widetilde{P}_{[4]}$, $[\mathbf{h}+2\mathbf{g}]\in \widetilde{P}_{[4]}$, $[\mathbf{h}+3\mathbf{g}]\in \widetilde{P}_{[4]}$.
\end{itemize}

\end{example}

By combining Theorems~\ref{thm:iff} and~\ref{thm:locality02}, we can provide a construction of $p$-ary alphabet-optimal $(2,p-1)$-LRCs which are Griesmer codes.

\begin{theorem}\label{thm:koptimal}
Let $p\geq 3$ be an odd prime, $\ell\geq 1$ and $m\geq 2$ be integers. Assume $\mathcal{A}_1,\mathcal{A}_2,\dots,\mathcal{A}_{\ell}$ are mutually disjoint subsets of $[m]$ and $M(\mathcal{A}_1,\mathcal{A}_2,\dots,\mathcal{A}_{\ell}) \leq  p-1$,
let $\mathcal{D}= P_{[m]} \setminus  (\bigcup_{i=1}^{\ell}P_{\mathcal{A}_i})$. Then $\mathcal{C}_{\mathcal{D}}$ constructed by \eqref{eq:framework} is an alphabet-optimal $(2,p-1)$-LRC with parameters
$$\left[\frac{p^m-\sum_{i=1}^{\ell}p^{|\mathcal{A}_i|}+\ell-1}{p-1} ,m, p^{m-1} -\sum_{i=1}^{\ell}p^{|\mathcal{A}_i|-1}\right].$$
\end{theorem}

\pf
Since $\mathcal{A}_1,\mathcal{A}_2,\dots,\mathcal{A}_{\ell}$ are mutually disjoint
and $M(\mathcal{A}_1,\mathcal{A}_2,\dots,\mathcal{A}_{\ell}) \leq  p-1$, we know that
there is at most one subset among $\{\mathcal{A}_i\}_{i=1}^{\ell}$, say $\mathcal{A}_j$, has size $m-1$, which means that there is at least a subset $\mathcal{A}^{*}\subset [m]$ with size $m-1$ such that $\mathcal{A}^{*}\neq \mathcal{A}_{i}$ for all $i\in [\ell]$. From Theorems~\ref{thm:iff} and~\ref{thm:locality02}, $\mathcal{C}_{\mathcal{D}}$ is a $p$-ary $[n,k,d]$ Griesmer code with $(2,p-1)$-locality, where $n=\frac{p^m-\sum_{i=1}^{\ell}p^{|\mathcal{A}_i|}+\ell-1}{p-1}$, $k=m$, $d=p^{m-1} -\sum_{i=1}^{\ell}p^{|\mathcal{A}_i|-1}$.

Since $\mathcal{A}_1,\mathcal{A}_2,\dots,\mathcal{A}_{\ell}$ are mutually disjoint, we have  $\sum_{i=1}^{\ell}p^{|\mathcal{A}_i|-1}\leq p^{m-2}+1$, then
$$\left\lceil\frac{p^{m-1}-\sum_{i=1}^{\ell}p^{|\mathcal{A}_i|-1}}{p^{m-2}}\right\rceil\geq p-1.$$
Note that
$$n=\sum_{j=0}^{m-1}\left\lceil\frac{p^{m-1}-\sum_{i=1}^{\ell}p^{|\mathcal{A}_i|-1}}{p^j}\right\rceil.$$
Thus
$$\sum_{j=0}^{m-2}\left\lceil\frac{p^{m-1}-\sum_{i=1}^{\ell}p^{|\mathcal{A}_i|-1}}{p^j}\right\rceil=  n-1,
 \sum_{j=0}^{m-3}\left\lceil\frac{p^{m-1}-\sum_{i=1}^{\ell}p^{|\mathcal{A}_i|-1}}{p^j}\right\rceil\leq n-p.$$
 Thanks to the Griesmer bound, $k_{{\rm opt}}^{(p)}(n -p, d) = m -2$. Utilizing the bound of \eqref{eq:CM2} with
$t = 1$, we get that $k \leq 2 + k_{{\rm opt}}^{(p)}(n-p, d) = m$. Therefore, the linear code $\mathcal{C}_{\mathcal{D}}$ achieves the bound of \eqref{eq:CM2}.

\EOP

\begin{remark}
In \cite{Luo2022}, the authors showed that code $\mathcal{C}_{\mathcal{D}}$ constructed in Theorem~\ref{thm:koptimal} is an alphabet-optimal $2$-LRC, here we prove that $\mathcal{C}_{\mathcal{D}}$ is actually an alphabet-optimal $(2,p-1)$-LRC, our results are more concise.
\end{remark}

Next, we give a construction of alphabet-optimal $(2,p-1)$-LRCs which are not Griesmer codes.

\begin{theorem}\label{thm:no}
Let $p\geq 3$ be an odd prime and $m=3$. Let $\mathcal{A}_1=\{1,2\}$ and $\mathcal{A}_2=\{2,3\}$. Let $\mathcal{D}^{c} = P_{\mathcal{A}_1} \cup P_{\mathcal{A}_2}$ and $\mathcal{D}= P_{[m]} \setminus \mathcal{D}^{c}$, then $C_{\mathcal{D}}$ constructed by \eqref{eq:framework} is a $p$-ary alphabet-optimal $(2,p-1)$-LRC with parameters $[p^2-p ,3, p^2 - 2p]$.
\end{theorem}

\pf From Theorems~\ref{thm:construction} and~\ref{thm:locality02}, $\mathcal{C}_{\mathcal{D}}$ is a $p$-ary $[n,k,d]$ code with $(2,p-1)$-locality, where $n=p^2-p$, $k=3$, $d=p^2 - 2p$. Utilizing the bound of \eqref{eq:CM2} with
$t = 1$, we get that $k \leq 2 + k_{{\rm opt}}^{(p)}(p^2-2p, p^2-2p) = 3$. Therefore, the linear code $\mathcal{C}_{\mathcal{D}}$ achieves the bound of \eqref{eq:CM2}. \EOP
\bigskip

Below we give an example to illustrate Theorem~\ref{thm:no}.

\begin{example}
Let $p=5$, $m=3$. Assume that $\mathcal{A}_1=\{1,2\}$ and $\mathcal{A}_2=\{2,3\}$. Let $\mathcal{D}^{c} = P_{\mathcal{A}_1} \cup P_{\mathcal{A}_2}$ and $\mathcal{D}= P_{[3]} \setminus \mathcal{D}^{c}$, then $C_{\mathcal{D}}$ constructed by \eqref{eq:framework} is a $5$-ary linear code with a generator matrix
\[\setcounter{MaxMatrixCols}{20}
G=\begin{bmatrix}
1&1&1&1&1&1&1&1&1&1&1&1&1&1&1&1&1&1&1&1\\
1&1&1&1&2&2&2&2&3&3&3&3&4&4&4&4&0&0&0&0\\
1&2&3&4&1&2&3&4&1&2&3&4&1&2&3&4&1&2&3&4
\end{bmatrix}.
\]
By Magma software, we know the parameters of $C_{\mathcal{D}}$ are $[20, 3, 15]$.
By Theorem~\ref{thm:locality02}, we can partition the matrix $G$ into the following four submatrices
\[\setcounter{MaxMatrixCols}{20}
\begin{bmatrix}
1&1&1&1&1\\
1&0&2&3&4\\
1&1&1&1&1
\end{bmatrix},
\begin{bmatrix}
1&1&1&1&1\\
1&0&2&3&4\\
2&2&2&2&2
\end{bmatrix},
\begin{bmatrix}
1&1&1&1&1\\
1&0&2&3&4\\
3&3&3&3&3
\end{bmatrix},
\begin{bmatrix}
1&1&1&1&1\\
1&0&2&3&4\\
4&4&4&4&4
\end{bmatrix}.
\]
In each submatrix, any two columns can be linearly combined to get the remaining three columns, from Definition~\ref{def:rdelta}, $C_{\mathcal{D}}$ is a $(2,4)$-LRC.
\end{example}

\begin{remark}
It is easy to check that the codes $C_{\mathcal{D}}$ constructed in Theorem~\ref{thm:no} are also Singleton-optimal.
This phenomenon reminds us that it may be an interesting topic to construct LRCs that achieve both Singleton-optimality and alphabet-optimality.
\end{remark}

\subsection{$(2,p)$-LRCs }
In Theorems~\ref{thm:locality} and~\ref{thm:locality02}, we utilize the inherent structure of defining sets to establish the $(2,p-2)$ and $(2,p-1)$-localities of $p$-ary linear code $\mathcal{C}_{\mathcal{D}}$, respectively. Now, we proceed to present a theorem that allows us to determine the $(2,p)$-locality of a $p$-ary linear code, where the defining set of this code is a subset of $P_{[m]}$, solely based on the cardinality of its defining set.
\begin{theorem}\label{thm:2PLOCALITY}
Let $p\geq2$ be a prime, $m > 2$ an integer. Suppose that $\mathcal{D}$ is a subset of $P_{[m]}$, $\mathcal{D}^{c}= P_{[m]}\setminus \mathcal{D}$. If $|\mathcal{D}^{c}|< \frac{p^{m-1}-1}{p-1}$, then
$\mathcal{C}_{\mathcal{D}}$ defined as in \eqref{eq:framework} is a $p$-ary $(2,p)$-LRC .

\end{theorem}
\pf
Let $\bF_p=\{0,-1,\alpha_1,\alpha_2,\dots, \alpha_{p-2}\}$.
We will show that for any nonzero $\mathbf{g} \in \mathcal{D}$, there always exists a $(p+1)$-size set $[\mathbf{g},\mathbf{h}]:=\{\mathbf{g},\mathbf{h},[\mathbf{h}-\mathbf{g}],[\mathbf{h}+\alpha_1\mathbf{g}],\dots, [\mathbf{h}+\alpha_{p-2}\mathbf{g}]\}\subset \mathcal{D}$ for some $\mathbf{h} \in \mathcal{D}\setminus \{\mathbf{g}\}$.

As any $2$ elements of the set $[\mathbf{g},\mathbf{h}]\setminus \{\mathbf{g}\}$ could be linearly combined to get $\mathbf{g}$, we call $[\mathbf{g},\mathbf{h}]\setminus \{\mathbf{g}\}$ a repair set of $\mathbf{g}$.
For different $\mathbf{h}_i, \mathbf{h}_j\in P_{[m]}\setminus \{\mathbf{g}\}$, it is easy to examine that $[\mathbf{g},\mathbf{h}_i]\cap [\mathbf{g},\mathbf{h}_j]=\{\mathbf{g}\}$.
So, there are $\frac{\frac{p^m-1}{p-1}-1}{p}=\frac{p^{m-1}-1}{p-1}$ disjoint repair sets of $\mathbf{g}$ in $P_{[m]}$.
Since $|\mathcal{D}^{c}| < \frac{p^{m-1}-1}{p-1}$, we have $|\mathcal{D}|=|P_{[m]}|-|\mathcal{D}^{c}|> p^{m-1}> (p-1)\frac{p^{m-1}-1}{p-1}$.
According to the Pigeonhole principle, for any vector $\mathbf{g}\in \mathcal{D}$, there always exists a repair set $[\mathbf{g},\mathbf{h}_0]\setminus \{\mathbf{g}\} \subset \mathcal{D}$, where $\mathbf{h}_0\in \mathcal{D}$ and $\mathbf{h}_0\neq \mathbf{g}$, which is equivalent to say that the coordinate occupied by $\mathbf{g}$ has $(2,p)$-locality. Since the chosen of $\mathbf{g}$ is arbitrary, all the coordinates of $\C_{\mathcal{D}}$ has $(2,p)$-locality.

\EOP

By combining Theorems~\ref{thm:iff} and~\ref{thm:2PLOCALITY}, we can provide a construction of $p$-ary alphabet-optimal $(2,p)$-LRCs.

\begin{theorem}\label{thm:koptimal2P}
Let $p\geq 3$ be an odd prime, $\ell\geq 1$ and $m\geq 2$ be integers. If $\mathcal{A}_1,\mathcal{A}_2, \dots ,\mathcal{A}_{\ell}$ are nonempty subsets
of $[m]$ satisfying
\begin{description}
  \item[\textnormal{(i)}] $\mathcal{A}_1,\mathcal{A}_2, \dots ,\mathcal{A}_{\ell}$ are mutually disjoint,
  \item[\textnormal{(ii)}] $M(\mathcal{A}_1,\mathcal{A}_2,\dots,\mathcal{A}_{\ell}) \leq  p-1$, and
  \item[\textnormal{(iii)}] $p^{m-1}>\sum_{i=1}^{\ell}p^{|\mathcal{A}_{i}|}$.
\end{description}
If $\mathcal{D}= P_{[m]} \setminus  (\bigcup_{i=1}^{\ell}P_{\mathcal{A}_i})$, $\mathcal{D}^{c}= P_{[m]}\setminus \mathcal{D}$, then $\mathcal{C}_{\mathcal{D}}$ constructed by \eqref{eq:framework} is an alphabet-optimal $(2,p)$-LRC with parameters
$$\left[\frac{p^m-\sum_{i=1}^{\ell}p^{|\mathcal{A}_i|}+\ell-1}{p-1} ,m, p^{m-1} -\sum_{i=1}^{\ell}p^{|\mathcal{A}_i|-1}\right].$$
\end{theorem}

\pf
From (i) and (iii), $|\mathcal{D}^{c}|=\frac{\sum_{i=1}^{\ell}(p^{|\mathcal{A}_i|}-1)}{p-1}< \frac{p^{m-1}-1}{p-1}$.
By Theorem~\ref{thm:2PLOCALITY}, $\mathcal{C}_{\mathcal{D}}$ has $(2,p)$-locality.
From (i)-(ii) and Theorem~\ref{thm:iff}, $\mathcal{C}_{\mathcal{D}}$ is a $p$-ary $[n,k,d]$ Griesmer code, where $n=\frac{p^m-\sum_{i=1}^{\ell}p^{|\mathcal{A}_i|}+\ell-1}{p-1}$, $k=m$, $d=p^{m-1} -\sum_{i=1}^{\ell}p^{|\mathcal{A}_i|-1}$.

From (iii), we have
$$\left\lceil\frac{p^{m-1}-\sum_{i=1}^{\ell}p^{|\mathcal{A}_i|-1}}{p^{m-2}}\right\rceil=p.$$

Note that
$$n=\sum_{j=0}^{m-1}\left\lceil\frac{p^{m-1}-\sum_{i=1}^{\ell}p^{|\mathcal{A}_i|-1}}{p^j}\right\rceil.$$
Thus
$$\sum_{j=0}^{m-2}\left\lceil\frac{p^{m-1}-\sum_{i=1}^{\ell}p^{|\mathcal{A}_i|-1}}{p^j}\right\rceil=n-1,
 \sum_{j=0}^{m-3}\left\lceil\frac{p^{m-1}-\sum_{i=1}^{\ell}p^{|\mathcal{A}_i|-1}}{p^j}\right\rceil= n-(p+1).$$
 Thanks to the Griesmer bound, $k_{{\rm opt}}^{(p)}(n -(p+1), d) = m -2$. Utilizing the bound of \eqref{eq:CM2} with
$t = 1$, we get that $k \leq 2 + k_{{\rm opt}}^{(p)}(n-(p+1), d) = m$. Therefore, the linear code $\mathcal{C}_{\mathcal{D}}$ achieves the bound of \eqref{eq:CM2}.

\EOP

\begin{remark}\label{rem:xiang}
From the proofs of Theorems~\ref{thm:locality}, \ref{thm:locality02} and \ref{thm:2PLOCALITY}, we can see that if we replace prime $p$ with any prime power $q$, the statements of localities for $p$-ary codes can be generalized to $q$-ary codes without any difficulties.
\end{remark}

\begin{corollary}
  Simplex codes over any finite field $\mathbb{F}_{q}$ are alphabet-optimal $(2,q)$-LRCs with respect to the bound \eqref{eq:CM2}.
\end{corollary}

\pf
From Remark~\ref{rem:xiang}, we know that any $q$-ary Simplex code $\mathcal{S}_m$  has $(2,q)$-locality. The parameters of $\mathcal{S}_m$ are $[n=\frac{q^m-1}{q-1},k=m,d=q^{m-1}]$. Utilizing the bound \eqref{eq:CM2} with $t = 1$, we get that $k \leq 2 + k_{{\rm opt}}^{(q)}(\frac{q^m-1}{q-1}-(q+1), q^{m-1}) \overset{(b)}{\leq} m$, where $(b)$ is from the Plotkin bound. The proof is done.

\EOP

\section{Alphabet-optimal $(r,\delta)$-LRCs with availability}

In this section, we will investigate the locality of the codes $\mathcal{C}_{(\mathcal{D}_1,\mathcal{D}_2,\dots,\mathcal{D}_s)}$ constructed in Theorem~\ref{thm:construction} with $s\geq 1$.
In the absence of ambiguity, we call an $[n,k,d]_q$ code alphabet-optimal if it achieves some upper bound for $k$ which takes the alphabet size $q$ into consideration.

As we can see in Section~\ref{sec:locality}, the set $\widetilde{P}_{[m]}$ is the beacon of proofs of $(r,\delta)$-localities. When $s> 1$, there are $s$ copies of $\widetilde{P}_{[m]}$ in the generator matrix of code $\mathcal{C}_{(\mathcal{D}_1,\mathcal{D}_2,\dots,\mathcal{D}_s)}$. Consequently, each coordinate of $\mathcal{C}_{(\mathcal{D}_1,\mathcal{D}_2,\dots,\mathcal{D}_s)}$ may possess $s$ disjoint repair sets. Next, we will give a formal definition of $(r,\delta)$-locality with availability from the aspect of generator matrix.

\begin{definition}\label{def:rt}
Let $\mathcal{C}$ be a $p$-ary linear code with generator matrix
$G = [\mathbf{g}_1, \dots, \mathbf{g}_n]$.
The $i$-th coordinate, $1 \leq i \leq n$, of $\mathcal{C}$ is said to have $(r, \delta)_{t}$-locality if there exist $t$ pairwise disjoint sets $\mathcal{I}^{(i)}_1 ,\dots, \mathcal{I}^{(i)}_{t}$, which are subsets of $ \{\mathbf{g}_1, \dots, \mathbf{g}_n\}\setminus \{\mathbf{g}_i\}$, satisfying that for each $j \in [t]$,
\begin{itemize}
  \item[(1)] $|\mathcal{I}^{(i)}_{j}\cup \{\mathbf{g}_i\}| \leq r+\delta-1$,
  \item[(2)] any $\delta-1$ vectors in $\mathcal{I}^{(i)}_{j}\cup \{\mathbf{g}_i\}$ are  linear combinations of the remaining vectors in $\mathcal{I}_{j}^{(i)}\cup \{\mathbf{g}_i\}$.
\end{itemize}
If all the coordinates of $\mathcal{C}$ have $(r,\delta)_{t}$-locality, then $\mathcal{C}$ is called an $(r,\delta)_{t}$-locally repairable code, or in short, $(r,\delta)_{t}$-LRC.
\end{definition}

\begin{remark}\label{rem:RTORT}
From the above definition, a $p$-ary $[n,k,d]$ code with $(r, \delta)_{t}$-locality is also a code with $(r, \delta)_{i}$-locality, $1\leq i \leq t-1$. So, for a linear code with $(r, \delta)_{t}$-locality, if it is $(r, \delta)$-alphabet-optimal, then it is also $(r, \delta)_{t}$-alphabet-optimal.
\end{remark}

From Remark~\ref{rem:RTORT}, for an $(r, \delta)_{t}$-LRC, where $t\geq 1$, we can prove that it is $(r, \delta)_{t}$-alphabet-optimal by proving that it is $(r, \delta)$-alphabet-optimal.

\begin{theorem}\label{thm:rtOLRC}
Let the notation be the same as in Theorem~\ref{thm:construction}. If $\mathcal{B}_1^{(j)},\mathcal{B}_2^{(j)},\dots,\mathcal{B}_{\ell_{j}}^{(j)}$ are mutually disjoint for every $j\in [s]$, $M(\mathcal{A}_1,\mathcal{A}_2,\dots,\mathcal{A}_{\ell}) \leq  p-1$,
and
\begin{equation}\label{eq:rt}
\left\lceil\frac{sp^{m-1}-\sum_{i=1}^{\ell}p^{|\mathcal{A}_i|-1}}{p^{m-1}}\right\rceil< p,
\end{equation}
then $\mathcal{C}_{(\mathcal{D}_1,\mathcal{D}_2,\dots,\mathcal{D}_s)}$ defined by \eqref{eq:define} is an alphabet-optimal $(2,p-1)_{s}$-LRC with parameters $$[\frac{sp^m-\sum_{i=1}^{\ell}p^{|\mathcal{A}_i|}+\ell-s}{p-1} ,m, sp^{m-1} -\sum_{i=1}^{\ell}p^{|\mathcal{A}_i|-1}].$$

\end{theorem}

\pf
From Theorem~\ref{thm:iff}, $\mathcal{C}_{(\mathcal{D}_1,\mathcal{D}_2,\dots,\mathcal{D}_s)}$ defined by \eqref{eq:define} is a $p$-ary $[n,k,d]$ Griesmer code, where $n=\frac{sp^m-\sum_{i=1}^{\ell}p^{|\mathcal{A}_i|}+\ell-s}{p-1}$, $k=m$, $d=sp^{m-1} -\sum_{i=1}^{\ell}p^{|\mathcal{A}_i|-1}$.

It is evident that all the $\mathcal{B}_{j}^{(i)}$, where $1\leq j\leq \ell_i$ and $1\leq i\leq s$, are proper subsets of $[m]$, so $\widetilde{P}_{[m]}\subset \mathcal{D}_{r}$ for all $1\leq r\leq s$. Then from Lemma~\ref{lem:vectorexist} and Definition~\ref{def:rt}, coordinates occupied by $\mathcal{D}_{i}\setminus \widetilde{P}_{[m]}$ have $(2,p-1)_s$-locality, where $1\leq i\leq s$.

From the proof of Theorem~\ref{thm:koptimal}, for each $i\in [s]$, there exists an $\mathcal{A}^{(i)}\subset [m]$ with size $m-1$ such that $\mathcal{A}^{(i)}\neq \mathcal{B}^{(i)}_{j}$ for all $j\in [\ell_{i}]$.
Then from Theorem~\ref{thm:locality02}, for any $\mathbf{g}$ in $\widetilde{P}_{[m]}$, we can find a $\mathbf{h}_i\in P_{\mathcal{A}^{(i)}}\subset \mathcal{D}_{i}\setminus \widetilde{P}_{[m]}$ such that there is a $p$-size set $\{\mathbf{g},\mathbf{h}_{i},\mathbf{h}_{i}+\alpha_1\mathbf{g},\dots,\mathbf{h}_{i}+\alpha_{p-2}\mathbf{g}\}\subset \mathcal{D}_{i}$ for all $1\leq i\leq s$, where $\mathbb{F}_{p}=\{0,-1,\alpha_1,\alpha_2,\dots,\alpha_{p-2}\}$. From Theorem~\ref{thm:koptimal} and Definition~\ref{def:rt}, coordinates occupied by $\widetilde{P}_{[m]}$ have $(2,p-1)_s$-locality.

In summary, $\mathcal{C}_{(\mathcal{D}_1,\mathcal{D}_2,\dots,\mathcal{D}_s)}$ is a $(2,p-1)_{s}$-LRC, and of course a $(2,p-1)$-LRC.

Note that
$$n=\sum_{j=0}^{m-1}\left\lceil\frac{sp^{m-1}-\sum_{i=1}^{\ell}p^{|\mathcal{A}_i|-1}}{p^j}\right\rceil,$$
from \eqref{eq:rt},
 we have
$$\sum_{j=0}^{m-2}\left\lceil\frac{sp^{m-1}-\sum_{i=1}^{\ell}p^{|\mathcal{A}_i|-1}}{p^j}\right\rceil>n-p.$$

Since $\mathcal{B}_1^{(j)},\mathcal{B}_2^{(j)},\dots,\mathcal{B}_{\ell_{j}}^{(j)}$ are mutually disjoint for every $j\in [s]$, we can deduce that
$$\left\lceil\frac{sp^{m-1}-\sum_{i=1}^{\ell}p^{|\mathcal{A}_i|-1}}{p^{m-2}}\right\rceil\geq s(p-1),$$
so
$$\sum_{j=0}^{m-3}\left\lceil\frac{p^{m-1}-\sum_{i=1}^{\ell}p^{|\mathcal{A}_i|-1}}{p^j}\right\rceil\leq n-1-s(p-1)\leq n-p.$$

 Thanks to the Griesmer bound, $k_{{\rm opt}}^{(p)}(n -p, d) = m -2$. Utilizing the bound \eqref{eq:CM2} with
$t = 1$, we can derive that $k \leq 2 + k_{{\rm opt}}^{(p)}(n-p, d) = m$, which means that $\mathcal{C}_{(\mathcal{D}_1,\mathcal{D}_2,\dots,\mathcal{D}_s)}$ is alphabet-optimal with respect to $(2,p-1)$-locality. From Remark~\ref{rem:RTORT}, $\mathcal{C}_{(\mathcal{D}_1,\mathcal{D}_2,\dots,\mathcal{D}_s)}$ is also alphabet-optimal with respect to $(2,p-1)_{s}$-locality.

\EOP

\begin{remark}
From the above theorem, the key of constructing alphabet-optimal $(r, \delta)_{s}$-LRCs is to make \eqref{eq:rt} hold, which can be fulfilled  when $s$ is small, for example, let $s< p$.
\end{remark}

\section{Conclusion}

In this paper, we first proposed a construction of linear codes $\mathcal{C}_{(\mathcal{D}_1,\dots\mathcal{D}_s)}$ over $\mathbb{F}_p$ by generalizing the constructions in \cite{Luo2022}. Similarly with \cite{Luo2022}, a necessary and sufficient condition for the linear codes $\mathcal{C}_{(\mathcal{D}_1,\dots\mathcal{D}_s)}$ to be Griesmer codes, a sufficient condition for $\mathcal{C}_{(\mathcal{D}_1,\dots\mathcal{D}_s)}$ to be distance-optimal, were presented. From which, some new constructions of Griesmer codes and distance-optimal codes can be derived.
Secondly, we proposed criteria for determining the $(2,p-2)$, $(2,p-1)$, and $(2,p)$-localities of $p$-ary linear codes constructed by eliminating elements from a complete projective space, and some alphabet-optimal $(2,p-1)$-LRCs and $(2,p)$-LRCs were provided. Specially, by showing that the methods of determining $(r,\delta)$-localities of $p$-ary code can be generalized to $q$-ary codes for any prime power $p$, we proved that the $q$-ary Simplex codes are alphabet-optimal $(2,q)$-LRCs. Finally, we explored the availability of $(r,\delta)$-LRCs constructed from the generalized framework \eqref{eq:define} with an alphabet-optimal construction. In the following research work, we
plan to explore more  $(r,\delta)$-localities of linear codes constructed from \eqref{eq:framework} and \eqref{eq:define} and to propose more alphabet-optimal $(r,\delta)$-LRCs.

%% Non-BibTeX users please use
%\begin{thebibliography}{}
%%
%% and use \bibitem to create references. Consult the Instructions
%% for authors for reference list style.
%%
%%\bibitem{RefJ}
%% Format for Journal Reference
%%Author, Article title, Journal, Volume, page numbers (year)
%% Format for books
%%\bibitem{RefB}
%%Author, Book title, page numbers. Publisher, place (year)
%
%
%
%
%
%
%\bibitem{1}
%Calderbank A.R., Rains E.M., Shor P.W., Sloane N.J.A.: Quantum error correction via codes over GF(4).
%IEEE Trans. Inf. Theory \mathbf{44}(4), 1369-1387 (1998).
%
%\bibitem{2}
%Ashikhmin A., Knill E.: Nonbinary quantum stabilizer codes. IEEE Trans. Inf. Theory \mathbf{47}(7), 3065-3072(2001).

\end{document}